\documentclass[twocolumn,10pt]{article}
\usepackage[top=2cm, bottom=2cm, left=1.8cm, right=1.8cm]{geometry}
\usepackage{times}

\usepackage{bbold}
\usepackage{amsmath}
\usepackage{algorithm}
\usepackage{algpseudocode}
\usepackage[dvipsnames]{xcolor}
\usepackage[hyphens]{url}
\usepackage{paralist}
\usepackage{graphicx}		%
	\setkeys{Gin}{width=\textwidth, height=\textheight, keepaspectratio}

\usepackage{booktabs}
\usepackage[square,sort&compress,comma,numbers]{natbib}
\usepackage[labelfont=bf]{caption}
\usepackage{amssymb}
\usepackage{pifont}
\usepackage{multirow}
\usepackage{tablefootnote}

\usepackage[bookmarks=true, bookmarksnumbered=true, colorlinks=true, linkcolor=linkcol, citecolor=citecol, urlcolor=urlcol, hypertexnames=true]{hyperref}

\definecolor{darkgreen}{RGB}{0, 100, 0}
\definecolor{linkcol}{rgb}{0.3,0,0}
\definecolor{citecol}{rgb}{0.3,0,0}
\definecolor{urlcol}{rgb}{0.3,0,0}
\definecolor{vlightgray}{gray}{0.925}

\let\OLDthebibliography\thebibliography
\renewcommand\thebibliography[1]{
  \OLDthebibliography{#1}
  \setlength{\parskip}{0pt}
  \setlength{\itemsep}{1pt plus 0.2ex}
}

\usepackage[hang,flushmargin]{footmisc}
\renewcommand{\footnotesize}{\fontsize{8}{9}\selectfont}

\makeatletter
\def\url@leostyle{%
  \@ifundefined{selectfont}{\def\UrlFont{}}%
  {\def\UrlFont{}}%
}
\makeatother
\usepackage[hyphenbreaks]{breakurl}

\newif\ifcomment
\commenttrue
\ifcomment
	\newcommand{\edc}[1]{\textbf{\em\color{red}EDC: #1}}
	\newcommand{\gvg}[1]{\textbf{\em\color{blue}GG: #1}}
\else
	\newcommand\edc[1]{}
	\newcommand\gvg[1]{}
\fi
\newif\ifcomment
\commenttrue
\definecolor{darkgreen}{RGB}{0, 90, 0}
\definecolor{darkblue}{RGB}{0, 0, 130}

\newcommand{\descr}[1]{\smallskip\noindent\textbf{#1}}
\newcommand{\descrit}[1]{\smallskip\noindent\textbf{\em #1}}
\newcommand{\provider}[1]{\textbf{\color{providercol}#1}}
\newcommand{\adversary}[1]{\textbf{\color{adversarycol}#1}}
\newcommand{\ccomment}[1]{{\scriptsize\Comment{\em #1}}}

\definecolor{providercol}{HTML}{2b8a3e}
\definecolor{adversarycol}{HTML}{c92a2a}
\usepackage{hyperref}

\newcommand{\mywidth}{0.49}

\usepackage[raggedright,compact]{titlesec}
\titlespacing*{\section}{0pt}{*4}{4pt}
\titlespacing*{\subsection}{0pt}{*3}{3pt}
\titlespacing*{\subsubsection}{0pt}{*2}{2pt}

\usepackage[skip=0pt,font=scriptsize]{subcaption}

\usepackage{titling}
\title{\bf The Inadequacy of Similarity-based Privacy Metrics: \\Privacy Attacks against ``Truly Anonymous'' Synthetic Datasets\thanks{Published in the Proceedings of the 46th IEEE Symposium on Security \& Privacy (IEEE S\&P 2025). Please cite the S\&P version.}}

\date{}
\author{Georgi Ganev$^1$ and Emiliano De Cristofaro$^2$\\[1ex]
\normalsize $^1$UCL and SAS, $^2$UC Riverside\\
\normalsize georgi.ganev.16@ucl.ac.uk, emilianodc@cs.ucr.edu}

\begin{document}

\maketitle
\sloppy

\thispagestyle{empty}
\pagestyle{plain}

\begin{abstract}
Generative models producing synthetic data are meant to provide a privacy-friendly approach to releasing data.
However, their privacy guarantees are only considered robust when models satisfy Differential Privacy (DP).
Alas, this is not a ubiquitous standard, as many leading companies (and, in fact, research papers) use ad-hoc privacy metrics based on testing the statistical {\em similarity} between synthetic and real data.

In this paper, we examine the privacy metrics used in real-world synthetic data deployments and demonstrate their unreliability in several ways.
First, we provide counter-examples where severe privacy violations occur even if the privacy tests pass and instantiate accurate membership and attribute inference attacks with minimal cost.
We then introduce \emph{ReconSyn}, a reconstruction attack that generates multiple synthetic datasets that are considered private by the metrics but actually leak information unique to individual records.
We show that \emph{ReconSyn} recovers 78--100\% of the outliers in the train data with only black-box access to a single fitted generative model and the privacy metrics.
In the process, we show that applying DP only to the model does not mitigate this attack, as using privacy metrics breaks the end-to-end DP pipeline.
\end{abstract}

\section{Introduction}
\label{sec:intro}
Generating synthetic tabular data using machine learning algorithms has attracted growing interest from the research community~\citep{jordon2022synthetic}, regulatory bodies~\citep{ico2022privacy, fca2023synthetic}, %
government agencies~\citep{benedetto2018creation, nist2018differential, nist2020differential}, as well as investors%
~\citep{crunchbase2022synthetic, techcrunch2022the}. %
The intuition is to %
learn the probability distribution of the real data and create new synthetic records by sampling from the trained model.
This approach promises an unlimited drop-in replacement for releasing sensitive data, de-biasing, data augmentation, etc.
However, models trained without robust privacy guarantees can memorize individual data points~\citep{carlini2019secret, webster2019detecting}, which enables attacks like membership and attribute inference~\citep{hayes2019logan, annamalai2023linear, chen2020gan, stadler2022synthetic, houssiau2022tapas, van2023membership}.
The established framework to limit information leakage from trained models %
is Differential Privacy (DP)~\citep{dwork2006calibrating, dwork2014algorithmic}; in the context of synthetic data, generative models should be trained while satisfying DP end to end~\cite{zhang2017privbayes, jordon2018pate, mckenna2021winning}.
However, while several synthetic data providers claim their products adhere to regulations like GDPR, HIPAA, or CCPA (see Section~\ref{subsec:solution}), many of them use ad-hoc heuristics to demonstrate privacy empirically rather than DP.
Some combine DP with unperturbed heuristics, which breaks the end-to-end DP pipeline and ultimately negates its privacy guarantees, as our analysis confirms.
This is worrisome, as these products are typically trained on sensitive data and are already deployed in highly regulated environments, e.g., healthcare~\citep{hradec2022multipurpose}.
Many synthetic data solutions use ad-hoc heuristics based on {\em similarity}, i.e., how similar/close synthetic records are to their nearest neighbor in the train data. %
More precisely, if enough synthetic points are too close according to pre-configured statistical tests vs.~holdout test data, either they are filtered out or the whole dataset is discarded.
Otherwise, the synthetic data is considered safe.
In other words, synthetic data should be similar and representative of the train data but not too close. %
These heuristics must be applied to every synthetic data release, as they are the only privacy mechanism some companies use.

\descr{Technical Roadmap.}
In this paper, we assess the validity and robustness of these similarity-based heuristics.
We discuss fundamental issues limiting their reliability, presenting counter-examples whereby privacy violations occur even if all similarity tests pass.
We show that their broad vulnerabilities enable membership and attribute inference, presenting a simple attack, \emph{DifferenceAttack}, achieving perfect performance with just a handful of calls to the metrics.

We then focus on the more complex problem of reconstructing records used to train the generative model.
We introduce \emph{ReconSyn}, a black-box attack %
reconstructing train data from low-density regions under realistic assumptions -- i.e.,
the adversary can only access a {\em single} fitted generative model and the metrics.
\emph{ReconSyn} operates by generating synthetic datasets deemed private by the metrics, and it is agnostic to the generative approach, dataset type, use case, etc.
It includes two subattacks: 1) \emph{SampleAttack}, which can only passively reconstruct data records generated by the trained model, and 2) \emph{SearchAttack}, which can actively augment the data points obtained in the first subattack.

\descr{Experimental Evaluation.}
We demonstrate the effectiveness of our attacks vis-\`a-vis five popular generative models for tabular data (PrivBayes~\cite{zhang2017privbayes}, MST~\cite{mckenna2021winning}, DPGAN~\cite{xie2018differentially}, PATE-GAN~\cite{jordon2018pate}, CTGAN~\cite{xu2019modeling}) and five commonly used datasets (\emph{2d Gauss}, \emph{Adult Small}, \emph{Adult}, \emph{Census}, and \emph{MNIST}).
We show that \emph{ReconSyn} reconstructs 78\% to 100\% of the underrepresented train data records (or outliers) with perfect precision in all settings, primarily due to the privacy leakage from the metrics.
Some models are more vulnerable: attacking graphical models (PrivBayes, MST) requires fewer rounds (i.e., API calls to the model/metrics) to achieve similar results as GANs. %
We focus on outliers as they potentially correspond to the most at-risk individuals.
Moreover, reconstructing outliers is inherently harder as the most difficult-to-model records likely reside in these low-density regions (as we confirm in Appendix~\ref{app:all}).
Similarly, we focus on tabular data, as: 1) the heuristics (and all studied generative models) are specifically designed and deployed for the tabular domain, and 2) there is an increasing number of real-world deployments in this space~\cite{benedetto2018creation, nhs2021ae, ons2023synthesising}.

\descr{Countermeasures (or Lack Thereof).}
As mentioned, providing meaningful privacy guarantees would entail training the generative models while satisfying DP.
However, we also demonstrate that DP models are still vulnerable to our attacks if used in conjunction with unperturbed heuristics, as the leakage persists through the metrics (see Section~\ref{subsec:mit}).

The next step would be to `DP-fy' the metrics while still running statistical pass/fail tests.
However, as the metrics must be applied every time, this would %
quickly deplete the privacy budget, thus %
preventing unlimited data generation, one of the providers' main selling point.
Also, one could, in theory, impose a limit on the number of queries allowed for each user, but this would also prevent unlimited data generation.
Moreover, synthetic data platforms are typically deployed on the client's premises or private clouds~\citep{gretel2024deploy, tonic2024deploy, mostly2024deploy, hazy2024deploy}; thus, the provider often lacks direct access to monitor or limit the number of queries.
Similarly, disabling access to the metric scores would compromise the provider's transparency and explainability, preventing users from verifying compliance claims without tangible proof. %

\begin{figure}[t!]
	\vspace{-0.3cm}
	\centering
	\begin{subfigure}{0.82\linewidth}
		\includegraphics{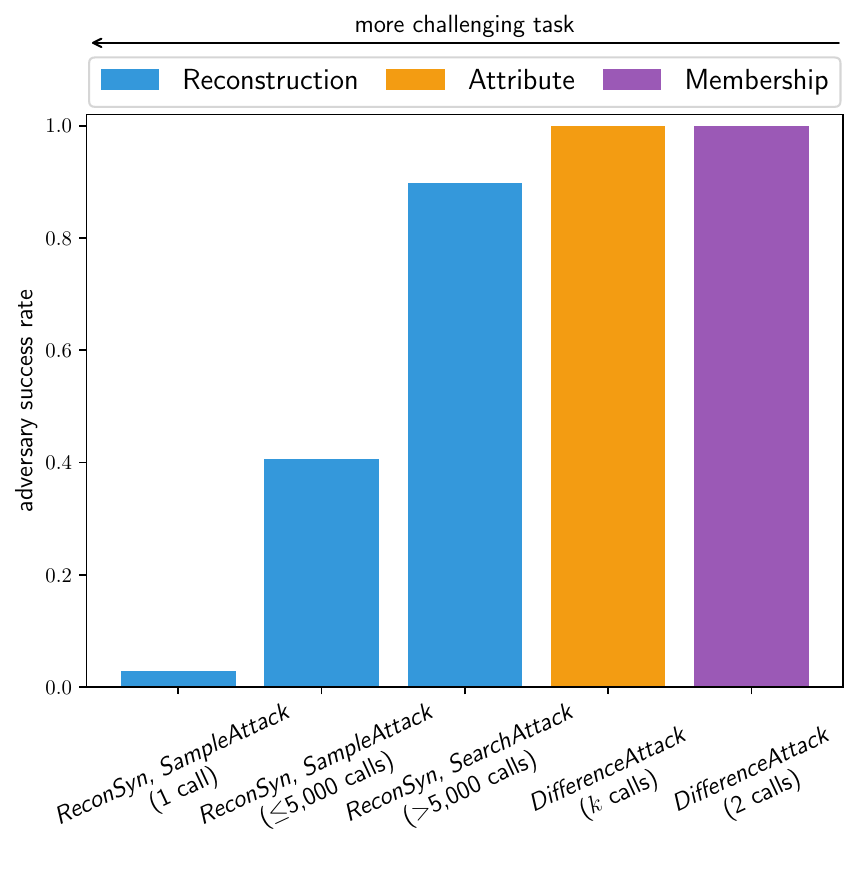}
	\end{subfigure}
	\caption{Overview of the  reconstruction success rate of \emph{ReconSyn} and \emph{DifferenceAttack}.
	\emph{ReconSyn} reconstructs outliers from the train data with success varying by attack phase (\emph{SampleAttack} and \emph{SearchAttack}) and the number of calls.	\emph{DifferenceAttack} achieves a 100\% success for membership and attribute inference ($k$ denotes the number of possible categories for the unknown attribute).}
	\label{fig:reconsyn_summary}
	\vspace{-0.2cm}
\end{figure}

\descr{Summary of Contributions:}\smallskip
\begin{compactenum}
	\item We analyze the undesirable properties of the most common privacy heuristics used in the wild to empirically measure and ``guarantee'' synthetic data privacy, demonstrating their inadequacy and failure at providing GDPR compliance.\smallskip
	\item We present a simple attack, \emph{DifferenceAttack}, which achieves perfect performance for membership and attribute inference with just a handful of metrics API calls; see the orange/purple bars in Figure~\ref{fig:reconsyn_summary}.\smallskip
	\item We introduce \emph{ReconSyn}, a novel reconstruction attack with minimal assumptions, i.e., black-box access to a single trained generative model and the privacy metrics.\smallskip
	After applying both attack phases, \emph{SampleAttack} and \emph{SearchAttack}, \emph{ReconSyn} reconstructs 78\%--100\% of the outliers in all datasets; see the blue bars in Figure~\ref{fig:reconsyn_summary}.\smallskip
	\item We stress the need to move away from reasoning about privacy in an ad-hoc, empirical way; rather, practitioners should adopt established end-to-end DP pipelines.
\end{compactenum}

\section{Preliminaries}
\label{sec:prelim}
In this section, we provide background information on synthetic data, Differential Privacy (DP), and generative models used in our evaluation.
We also introduce and review similarity-based privacy metrics used by leading companies.

\subsection{Synthetic Data and (DP) Generative Models}
\label{subsec:genmodels}

\descr{Synthetic Data from Generative Models.}
A sample dataset $D_{train}^n$ (consisting of $n$ iid records) %
is used as input to the generative model training algorithm $G(D_{train}^n)$ during the fitting step.
Next, $G(D_{train}^n)$ updates its parameters $\theta$ to capture a representation of $P(D_{train}^n)$ and outputs the trained model $G_{\overline{\theta}}(D_{train}^n)$.
Then, $G_{\overline{\theta}}(D_{train}^n)$ can be used to repeatedly sample a synthetic dataset of arbitrary size $n^{\prime}$, $D_{synth}^{n^{\prime}}{\sim}G_{\overline{\theta}}(D_{train}^n)$.
Finally, we use $D_{train}^{out} \in D_{train}^n$ to denote the outliers or train records with low density.
We refer to Appendix~\ref{app:data} for details on the datasets used in our experiments and how we define outliers for each of them.

\descr{Differential Privacy (DP)}
is a mathematical definition that formally bounds the probability of distinguishing whether any given individual's data was included in the input dataset by looking at the output (e.g., a trained model).
More formally, a randomized algorithm $\mathcal{A}$ satisfies ($\epsilon, \delta$)-DP if, for all possible outputs $S$, and all neighboring datasets $D$ and $D^{\prime}$ ($D$ and $D^{\prime}$ are identical except for a single individual's data), it holds that~\citep{dwork2006calibrating, dwork2014algorithmic}:
\begin{equation}
P[{\mathcal{A}}(D)\in S]\leq \exp \left(\epsilon \right)\cdot P[{\mathcal{A}}(D^{\prime})\in S] + \delta
\end{equation}
Note that $\epsilon$ (aka the privacy budget) is a positive, real number quantifying the information leakage, while $\delta$, usually an asymptotically small real number, allows for a probability of failure.
DP is generally achieved through noisy/random mechanisms that could be combined together, as the overall privacy budget can be tracked due to DP's {\em composition} property.
Also, DP-trained models can be re-used, as per the {\em post-processing} property, without further privacy leakage.

\descr{(DP) Generative Models.}
We focus on two types of generative approaches, graphical models (PrivBayes and MST) and GANs (DPGAN, PATE-GAN, and CTGAN), as they are generally considered to perform best in practice in the tabular domain~\citep{nist2018differential}.
All algorithms have open-source implementations and rely on different modeling techniques and, when applicable, DP mechanisms.
We also present two baseline models, Independent and Random.
Except for CTGAN, all support DP training.
Essentially, graphical models break down the joint distribution of the dataset to explicit lower-dimensional marginals, while GANs approximate the distribution implicitly by training two neural networks with opposing goals: a generator, which creates realistic synthetic data from noise, and a discriminator, which distinguishes real from synthetic data points.

\descrit{PrivBayes~\citep{zhang2017privbayes}}
follows a two-step process -- finding an optimal Bayesian network and estimating the resulting conditional distributions.
First, it builds the network by iteratively selecting a node that maximizes the mutual information between the already chosen parent nodes and one of the remaining candidate child nodes.
Second, it computes noisy distributions using the Laplace mechanism~\citep{dwork2006calibrating}.

\descrit{MST~\citep{mckenna2021winning}}
uses a similar approach.
First, it utilizes Private-PGM~\citep{mckenna2019graphical} (method inferring data distribution from noisy measurements) to form a maximum spanning tree of the underlying graph by selecting all one-way and subset of two-way marginals (attribute pairs).
Then, the marginals are measured privately using the Gaussian mechanism~\citep{dwork2006our}.

\descrit{DPGAN~\citep{xie2018differentially}}
modifies the GAN training procedure to satisfy DP.
It uses DP-SGD~\citep{abadi2016deep} to sanitize the discriminator's gradients (to clip the norm of individual ones and apply the Gaussian mechanism to the sum), which guarantees the privacy of the generator by the post-processing property.

\descrit{PATE-GAN~\citep{jordon2018pate}}
combines an adapted PATE framework~\citep{papernot2017semi, papernot2018scalable} with a GAN to train a generator, $t$ teacher-discriminators, and a student-discriminator.
The teacher-discriminators are presented with disjoint partitions of the data and are optimized to minimize their loss with respect to the generator.
The student-discriminator is trained on noisily aggregated labels provided by the teacher-discriminators while its gradients are sent to train the generator.

\descrit{CTGAN~\citep{xu2019modeling}}
is one of the most widely used non-DP generative models.
It uses mode-specific normalization to overcome the non-Gaussian and multimodal distribution of mixed-type tabular datasets.
It relies on a conditional generator and training-by-sampling to capture data imbalances.

\descrit{Independent \& Random}
are used as baselines.
Independent is a common baseline model for (DP) synthetic data generation~\citep{ping2017datasynthesizer, tao2022benchmarking, mahiou2022dpart}.
It models all columns independently, thus attempting to preserve the marginal distributions but omitting higher-order correlations.
We then create the Random model by randomly sampling from the distinct values from all columns (thus resulting in even lower utility).

\subsection{Commercial Synthetic Data Solutions}
\label{subsec:solution}

\begin{table}[t!]
	\small
	\centering
		\setlength{\tabcolsep}{2pt}
		\begin{tabular}{l@{}ccl}
			\toprule
				\bf Company																							& \bf Claimed				& \bf DP				& \bf Ad-hoc 	Privacy								 							  \\
																																& \bf Compliance		& \bf Models		& \bf  Heuristics																		\\
			\midrule
				Gretel$^1$~\citep{gretel2024gretel}											& \checkmark				& \checkmark		& SF, OF~\citep{gretel2021introducing}							\\
				Tonic~\citep{tonic2024tonic}	 													& \checkmark				& \checkmark		& DCR~\citep{tonic2023can}													\\
			 	Mostly AI~\citep{mostly2024mostly}											& \checkmark				& \checkmark$^2$ & IMS, DCR, NNDR~\citep{mostly2022mostly}					\\
				Hazy$^3$~\citep{hazy2024hazy}		 		 										& \checkmark				& \checkmark		& DCR~\citep{hazy2024privacy}												\\
				Aindo~\citep{aindo2024aindo}														&	\checkmark				& \ding{56}			& NNDR~\citep{panfilo2022generating, panfilo2023a}	\\
				DataCebo~\citep{datacebo2024datacebo} 									&	--								&	\ding{56}			& IMS~\citep{sdv2024synthetic} 											\\
				Syntegra~\citep{syntegra2024syntegra}										& \checkmark				& \ding{56}			& IMS, DCR~\citep{syntegra2021fidelity}							\\
				YData~\citep{ydata2024ydata}														& \checkmark				& \checkmark		& IMS, DCR~\citep{ydata2023how}											\\
			 	Synthesized~\citep{synthesized2024synthesized}					& \checkmark				& \checkmark		& SF~\citep{synthesized2024dstrict} 								\\
				Syntho~\citep{syntho2024syntho}													& \checkmark				& \ding{56}			& IMS, DCR, NNDR~\citep{syntho2024report}						\\
				Replica$^4$~\citep{replica2024replica}									& \checkmark				& \ding{56}			& SF~\citep{replica2020practical} 									\\
				Statice$^5$~\citep{statice2024statice} 									& \checkmark				& \checkmark		& DCR$^6$~\citep{statice2022anonymization}					\\
			\bottomrule
\multicolumn{4}{p{0.99\linewidth}}{\scriptsize $^1$Acquired by NVIDIA in Mar 2025.~$^2$DP support, alongside SBPMs, added in Nov 2024.~$^3$Acquired by SAS in Nov 2024.~$^4$Acquired by Aetion in Jan 2022.~$^5$Acquired by Anonos in Nov 2022.~$^6$Used for linkability and inference risk metrics~\citep{giomi2022unified}.}\\
			\bottomrule
		\end{tabular}
		\caption{List of representative synthetic data companies, whether they claim to be offering regulatory-compliant synthetic data, and the similarity-based privacy metrics/filters they use.}
		\label{tab:industry}
\end{table}

We have been tracking the main synthetic data providers (Gretel, Tonic, Mostly AI, Hazy, Aindo, DataCebo, Syntegra, YData, Synthesized, Syntho, Replica Analytics, and Statice) %
and examined publicly available information on their claims of regulatory compliance and their approach to privacy, i.e., whether they support DP training and what heuristics they use.
As summarized in Table~\ref{tab:industry}, they mainly use five heuristics, which we describe in Section~\ref{subsec:metrics}: three metrics, Identical Match Share~(IMS), Distance to Closest Records~(DCR), Nearest Neighbor Distance Ratio~(NNDR), and two filters, Similarity Filter~(SF) and Outlier Filter~(OF).
{\em N.B. For simplicity, we refer to the three metrics as similarity-based privacy metrics (SBPMs) and use the term (ad-hoc) heuristics to include all metrics and filters.} %

Almost all companies claim on their websites %
that their synthetic data products comply with regulations like GDPR, HIPAA, CCPA, etc., even though there are no established standards for mapping privacy to regulatory frameworks in the context of synthetic data.
We also note that %
larger and better-funded companies tend to report adopting DP.
This might not be surprising, as integrating DP in production requires specialized technical knowledge.
Five of the twelve companies do not support DP but claim to guarantee privacy through one or more of the three SBPMs, while two combine it with the two privacy filters.

Note that, besides commercial products, peer-reviewed research outputs have also used models that exclusively rely on the heuristics -- see, e.g.,~\citep{park2018data, lu2019empirical, zhao2021ctab, panfilo2022generating, venugopal2022privacy, borisov2023language, guillaudeux2023patient, solatorio2023realtabformer, liu2023tabular, yoon2023ehr, sivakumar2023generativemtd, kotelnikov2023tabddpm, zhang2023generative, damico2023synthetic, zhang2024mixedtype}.

\subsection{Similarity-based Privacy Metrics (SBPMs)}
\label{subsec:metrics}
We now review the five ad-hoc heuristics synthetic data companies use to provide privacy assurances.
As mentioned, these include three SBPMs used to measure the privacy of a synthetic dataset as a whole and run pass/fail statistical tests, and two filters used to remove records from the generated data based on their similarity to train records or outliers.

A common pre-processing step, which we will follow in our experiments, is to {\em discretize} the data.
The input to all the metrics is the train dataset ($D_{train}^n$), the synthetic data ($D_{synth}^{n^{\prime}}$), and the holdout test dataset, $D_{test}^n$.
$D_{train}^n$ and $D_{test}^n$ have the same size $n$ and come from the same distribution, but the latter is not used to train the model.
The idea behind SBPMs is that synthetic records should be as close as possible to train ones, but not too close, i.e., not closer than what would be expected from the holdout records~\citep{platzer2021holdout, mobey2022help}.
More precisely, SBPMs compute the closest pairwise distances (Hamming distance for discrete and Euclidean for continuous data) for ($D_{train}^n$, $D_{synth}^{n^{\prime}}$) and ($D_{train}^n$, $D_{test}^n$), compare their distributions, and run a pass/fail test.
The passing criterion is a comparison between simple statistics calculated from the two distributions, e.g., the average or the 5th percentile, while the output of the metrics is the pass/fail flag alongside the actual statistics~\citep{mostly2022mostly}.
If all three tests pass, the synthetic data is deemed {\em ``truly anonymous''}~\citep{mostly2020truly} and could be freely shared alongside the privacy scores.
Otherwise, the data is not %
to be released.

\descrit{Identical Match Share (IMS)}
is a privacy metric that captures the proportion of identical copies between train and synthetic records.
The test passes if that proportion is smaller or equal to the one between train and test datasets.
IMS is used or advocated by companies~\cite{mostly2020truly, syntegra2021fidelity, ydata2023how, sdv2024synthetic, syntho2024report}, and scientific papers/blogposts~\citep{lu2019empirical, aws2022how, sen2023diverse, damico2023synthetic}.

\descrit{Distance to Closest Records (DCR)}
also compares the two sets of distances.
It looks at the overall distribution of the distances to the nearest neighbors or closest records.
The test passes if ($D_{train}^n$, $D_{synth}^{n^{\prime}}$)-5th percentile is larger or equal than the other pair.
DCR is supposed to protect against settings where the train data is just slightly perturbed and presented as synthetic.
Several companies~\cite{tonic2023can, mostly2020truly, hazy2024privacy, syntegra2021fidelity, ydata2023how, statice2022anonymization, syntho2024report}, and scientific studies/blogposts~\citep{park2018data, lu2019empirical, zhao2021ctab, aws2022how, venugopal2022privacy, borisov2023language, kotelnikov2023tabddpm, guillaudeux2023patient, liu2023tabular, yoon2023ehr, sivakumar2023generativemtd, zhang2023generative, zhang2024mixedtype} use it.

\descrit{Nearest Neighbor Distance Ratio (NNDR)}
is very similar to DCR, but the nearest neighbors' distances are divided by the distance to the second nearest neighbor.
The idea is to add further protection for outliers by computing relative rather than absolute distances.
NNDR, too, compares the 5th percentile between the two sets of distributions and is used by a few companies~\cite{mostly2020truly, aindo2024aindo, syntho2024report} and in research papers~\citep{zhao2021ctab, panfilo2022generating, panfilo2023a, guillaudeux2023patient, sivakumar2023generativemtd, damico2023synthetic}.

\descrit{Similarity Filter (SF)}
is similar in spirit to the privacy metrics, but rather than just measuring similarity, it excludes or filters out individual synthetic data points if they are identical or too close to train ones.
Essentially, SF aims to ensure that no synthetic record is overly similar to a train one.
It is used by several companies~\cite{replica2020practical, gretel2021introducing, synthesized2024dstrict}.

\descrit{Outlier Filter (OF)}
focuses on the outliers; %
it removes synthetic records that could be considered outliers with respect to the train data, and is used by one company~\cite{gretel2021introducing}.

\descr{Passing Criteria.}
Throughout our experiments, unless stated otherwise, we will use the passing criteria from~\citep{mostly2020truly} -- i.e., a synthetic dataset (for whose generation none of the filters were used) is considered private if all three privacy tests pass (corresponding to IMS, DCR, and NNDR).

\section{Fundamental Issues of SBPMs}
\label{sec:issues}
In this section, we identify and discuss several fundamental issues of using Similarity-Based Privacy Metrics (SBPMs) to reason about privacy through pass/fail tests.

\descr{I1: No Theoretical Guarantees.}
First and foremost, SBPMs do not provide any theoretical or analytical guarantees.
Instead, they rely on arbitrarily chosen statistical tests, which prompts several questions, e.g., why choose these specific tests instead of others? %
How were the passing criteria selected?
Furthermore, SBPMs do not rule out vulnerabilities to current or future adversarial privacy attacks. %

\descr{I2: Strong Implicit Threat Model.}
While SBPMs do not formally define a threat model or a strategic adversary (thus, ignoring fundamental security principles~\citep{anderson2020security}), they implicitly allow for unlimited synthetic data querying without repercussions.
This is because they treat privacy as a binary property (see I3 below) and a property of the synthetic data rather than the generating process (see I5).

\descr{I3: Privacy as Binary Property.}
SBPMs treat privacy leakage as a binary property: the synthetic data is either ``truly'' private or not.
In fact, using pass/fail tests may remove analysts' %
ability to measure privacy leakage across a continuous interval.
This has two consequences.
First, it is hard to know what choices (e.g., models, hyperparameters, etc.) contribute to making the synthetic data private. %
Second, releasing a single private synthetic dataset is deemed as safe as releasing many (as long as they pass the tests), even though this increases leakage as the provider needs to query the train/test data every time new data is generated.
Alas, the Fundamental Law of Information Reconstruction~\citep{dwork2014algorithmic} tells us that ``overly accurate answers to too many questions will destroy privacy in a spectacular way.''

\begin{figure*}[t!]
	\vspace{-0.3cm}
	\begin{minipage}[t]{0.325\linewidth}
		\centering
		\includegraphics[width=\linewidth]{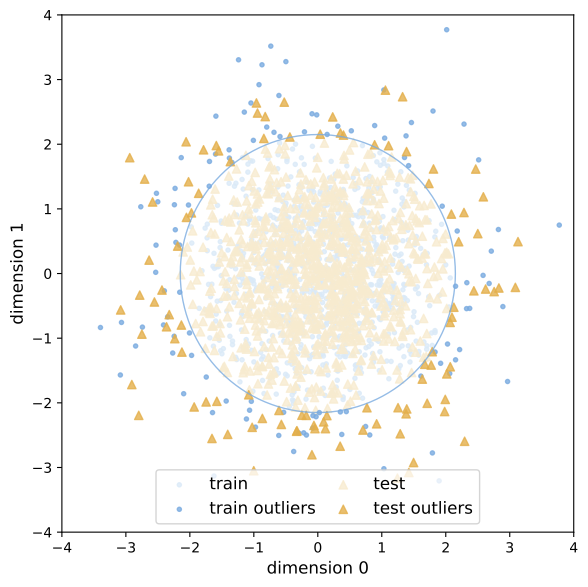}
		\caption{Train and test data, \emph{2d Gauss}.} %
		\label{fig:2d_gauss_fixed}
	\end{minipage}
\hfill
	\begin{minipage}[t]{0.32\linewidth}
		\centering
		\includegraphics[width=\linewidth]{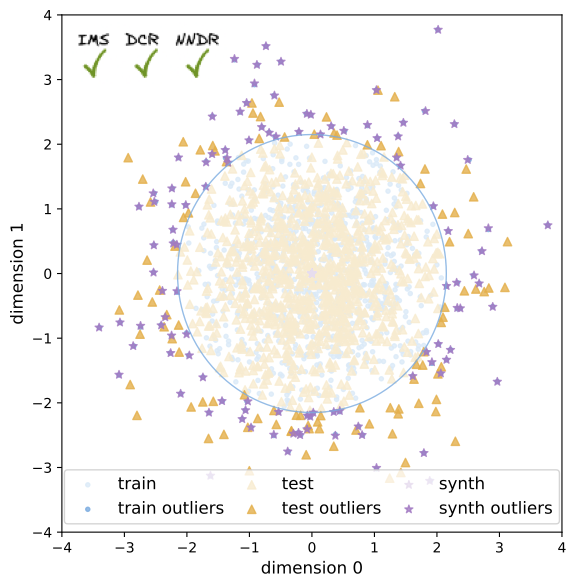}
		\caption{Synth data reproducing all train outliers, \emph{2d Gauss}.}
		\label{fig:wonky_outliers}
	\end{minipage}
\hfill
	\begin{minipage}[t]{0.325\linewidth}
		\centering
		\includegraphics[width=\linewidth]{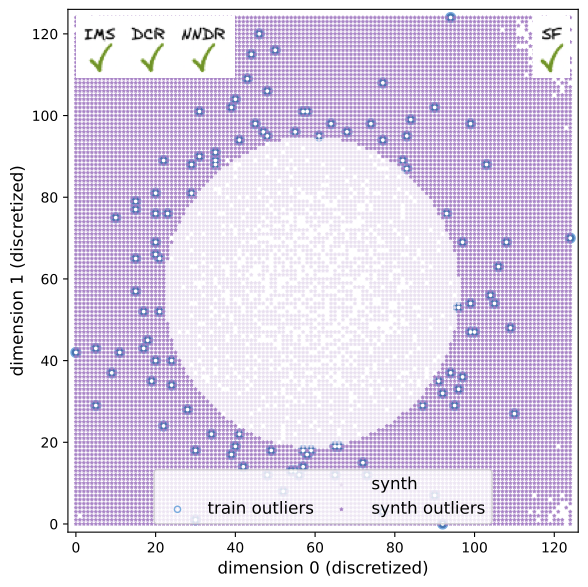}
		\caption{Applying SF resulting in ``Swiss cheese,'' \emph{2d Gauss}.}
		\label{fig:2d_sf_recon}
	\end{minipage}
	\vspace{-0.2cm}
\end{figure*}

\descr{I4: Non-Contrastive Process.}
SBPMs are computed in a non-contrastive way, i.e., they do not compare the computations when an individual is included or not.
Since there is no noise or randomness ingested into the process, plausible deniability is ruled out.
Thus, calculating the privacy metrics leads to vulnerability to a variety of attacks, including simple ones like differencing attacks.
For example, if an adversary makes two calls to the metrics, one with and one without a particular individual, they can deduce some information (e.g., whether the individual is an exact match or closer than 5th percentile) with 100\% confidence since the computations carry no uncertainty.

\descr{I5: Privacy as Data Property.}
SBPMs expect a single synthetic dataset as input, which has several implications.
First, it means we measure the privacy of a specific dataset and not the generative model/process.
In fact, the metrics are agnostic to how the data was generated, which also allows for manually altering the data.
Therefore, privacy becomes a property of the data rather than the generating process.
Also, SBPMs require running the metrics on each and every generated synthetic data in order to deem them private, which, unfortunately, actually leaks more privacy (as discussed in I3).
Second, the specific synthetic dataset may or may not be representative of the distribution captured by the model, which could lead to inconsistent results across generation runs.
Typically, privacy is defined as a statistical property over many such instances.

\descr{I6: Lack of Worst-Case Analysis.}
All SBPMs use simple statistics (average or 5th percentile) as passing criteria.
This leaves room for maliciously crafted synthetic datasets that might pass the tests but still reveal sensitive data.
Also, this does not protect against worst-case scenarios, i.e., memorization and replication of outliers, which, combined with the lack of plausible deniability (from I4), increases the adversary's chance of launching a successful attack.

Unfortunately, using a held-out dataset does not alleviate the problem due to the so-called ``Generalization Implies Privacy'' fallacy~\cite{del2023bounding}, %
i.e., privacy is a worst-case problem while generalization is average-case.
Put simply, even if all tests pass, i.e., the model generalizes, memorization cannot be ruled out~\citep{song2017machine}.

\smallskip
In Appendix~\ref{app:issues}, we discuss three additional fundamental issues with SMPMs, relating to incorrect interpretation, risk underestimation, and data access.

\section{Counter-Examples}
\label{sec:examples}
Next, we discuss three counter-examples %
whereby there is obvious privacy leakage even if all three tests from SBPMs pass (or any of the privacy filters are applied).
We use \emph{2d Gauss} dataset (see Appendix~\ref{app:data}), which we visualize over its two dimensions in Figure~\ref{fig:2d_gauss_fixed}.
Since all attributes are continuous, we use the Euclidean distance to make the computations more accurate.
In Appendix~\ref{app:examples}, we discuss three additional counter-examples, further showcasing the inconsistent nature of the metrics and filters.

\begin{figure*}[t!]
	\vspace{-0.3cm}
	\centering
	\includegraphics[width=0.99\textwidth]{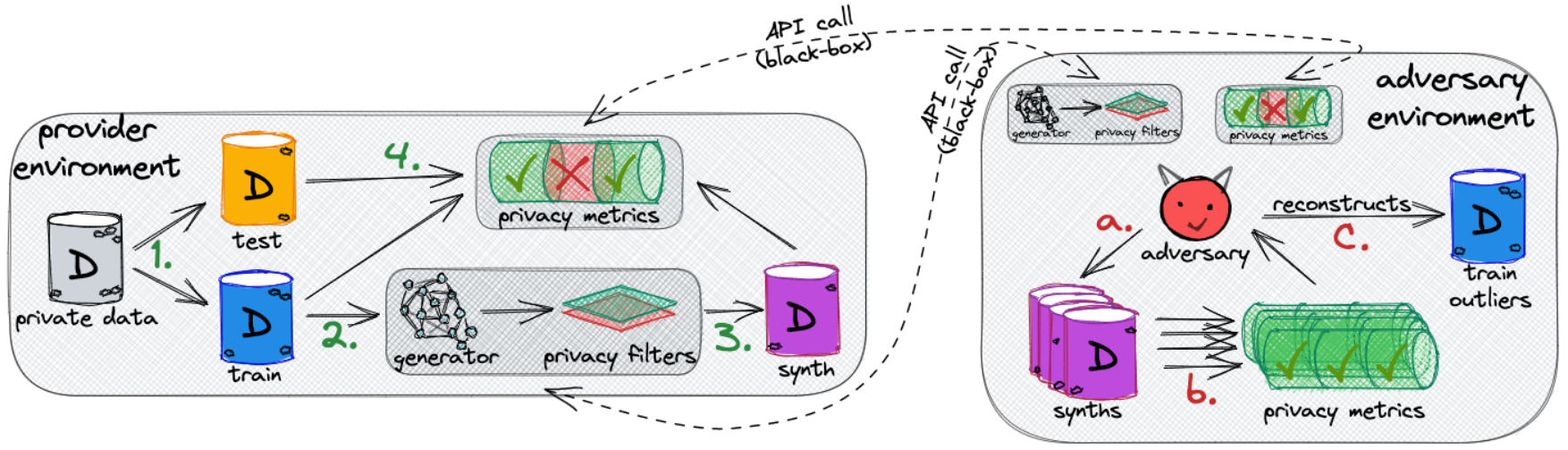}
	\caption{\emph{ReconSyn} Overview.
					 The provider \provider{1.} splits the private data into train/test, \provider{2.} fits the generative model on train data, \provider{3.} generates synthetic data (privacy filters are applied), \provider{4.} runs the privacy metrics on synthetic data.
					 The adversary can make API calls (i.e., black-box access) to the fitted generative model and privacy metrics.
					 They \adversary{a.} generate synthetic datasets, \adversary{b.} run them through the privacy metrics to observe the pass/fail tests and scores (if tests pass), \adversary{c.} reconstruct train data outliers (through \emph{SampleAttack} and \emph{SearchAttack}).}
	\label{fig:reconsyn}
	\vspace{-0.2cm}
\end{figure*}

\descr{\em CE1.~Leaking All Test Data.}
Assume a synthetic dataset that is an exact replica of the test data.
All privacy tests pass as the two distributions of distances, ($D_{train}^n$, $D_{synth}^{n^{\prime}}$) and ($D_{train}^n$, $D_{test}^n$), are identical.
Following the supposed guarantees provided by the metrics, we would be free to release this dataset.
Naturally, publishing half of the sensitive records cannot be considered privacy-preserving.

\descr{\em CE2.~Leaking All Train Outliers.}
Next, assume that the synthetic data contains all train outliers (with an indistinguishably small perturbation) and the value (0, 0) repeated five times the size of the train data, as displayed in Figure~\ref{fig:wonky_outliers}.
Again, all privacy tests pass: there are no exact matches, and even though the synthetic outliers are extremely close to the train ones, the large sample size of zeros skews the distances enough to fool both DCR and NNDR.
If this synthetic dataset is released, individuals whose data corresponds to the outliers and whose sensitive attributes are leaked would be unconvinced that their privacy is actually preserved~\citep{ons2018privacy}.

\descr{\em CE3.~SF \& Reconstruction.}
We can reconstruct {\em all} outliers if Similarity Filter (SF) is applied.
We assume access to an oracle possessing knowledge of the %
generative process.
Suppose we use the oracle to sample 100,000 synthetic datasets, apply SF, and select the datasets passing all privacy tests (we plot them in Figure~\ref{fig:2d_sf_recon}).
Since no generative model was trained (i.e., the train data was never exposed to a model), any data directly sampled from the oracle preserves privacy.

However, all train outliers can immediately be detected and reconstructed from the `holes' in the data---we refer to this emerging pattern as ``Swiss Cheese.''
This simple experiment shows that even though SF could naively be considered an additional privacy layer, filtering data out could actually {\em expose} data points.
Furthermore, outliers could uniquely be identified since they are typically in low-density regions.
A similar pattern could be observed if the Outlier Filter (OF) is applied.
This phenomenon is also discussed in~\citep{jordon2022synthetic}, although not demonstrated experimentally. %
In Appendix~\ref{app:examples}, we also discuss how this simple experiment fails to scale to higher dimensions unless specific records are targeted.

\smallskip
Given the severe privacy leakage coming from the two privacy filters, in the rest of the paper, we do not consider them and instead focus on the three SBPMs. %

\section{Attacks}
In this section, we present two privacy attacks exploiting the inadequacy of similarity-based privacy metrics (SBPMs).
We start with the simple(r) \emph{DifferenceAttack}, which is geared for membership and attribute inference, then present \emph{ReconSyn}, a (more complex) reconstruction attack.

\subsection{Adversarial Model}
\label{sec:model}
We consider a synthetic data provider that has access to train and test datasets, i.e., $D_{train}^n$ and $D_{test}^n$, trains a generative model $G_{\overline{\theta}}(D_{train}^n)$, and generates multiple synthetic datasets $\{D^{n^{\prime}}_{synth}\}_i$.
As discussed, the provider deems the synthetic datasets `private' if they pass all privacy tests from the three privacy metrics (IMS, DCR, and NNDR).

We assume a strategic adversary aiming to disprove the blanket statement that ``if the tests pass, the synthetic data is truly anonymous.''
More precisely, they build synthetic datasets deemed private by the provider and undertake the following adversarial tasks: 1) membership inference, i.e., determining whether a target record was part of the train data; 2) attribute inference, i.e., predicting an unknown attribute of a target record within the train data given the other attributes; 3) reconstructing the train data outliers ($D_{train}^{out}$).

\descr{Assumptions.}
We assume the adversary has black-box access to the trained generative model $G_{\overline{\theta}}(D_{train}^n)$ and the privacy metrics.
More precisely, we consider three adversarial assumptions that are consistent with industry practice.

\descrit{1) Black-box access to the generator.}
The ability to generate synthetic data from the model is a major selling point for adopting synthetic data in commercial applications~\citep{gretel2024gretel, mostly2024mostly, hazy2024hazy}.
State-of-the-art attacks against generative models also rely on generating (large) collections of datasets (e.g., up to 1,000,000 for tabular data~\citep{hilprecht2019monte} and 175,000,000 for images~\citep{carlini2023extracting}).
Also, black-box access is a standard, minimal assumption of privacy attacks against generative models~\citep{hayes2019logan, hilprecht2019monte, stadler2022synthetic, annamalai2023linear, nasr2023scalable} and in general against ML~\cite{papernot2018sok}.

\descrit{2) Adding/removing records to the synthetic data.}
The adversary can conditionally generate data or update the synthetic data with desirable patterns (e.g., bias correction, rebalancing, signal boosting) as this does not have an effect on the train data privacy as long as all privacy tests pass (see I2 in Section~\ref{sec:issues}).
Data augmentation/conditional generation is also a popular use case advertised by synthetic data companies~\citep{tonic2022how, gretel2024conditional, mostly2023what, replica2024replica}.

\descrit{3) Querying the privacy metrics}.
The adversary can run the privacy metrics and access the scores if all tests pass.
For companies that rely on empirical metrics, privacy is considered a property of the data (see I5 in Section~\ref{sec:issues}); thus, querying the metrics on each synthetic dataset and providing access to the scores to their users becomes essential to providing privacy assurances.
To be precise, this is {\em explicitly} offered by companies~\citep{mostly2022mostly, ydata2023how, syntho2024report}; basically, %
it is their only way to ``demonstrate'' privacy to users.
Prior work on attacking deployed systems also use (large) amounts of queries -- e.g., adversaries targeting large language models can query them up to 400,000,000 times to steal parts of the model~\citep{carlini2024stealing}, generate up to 20,000,000,000 tokens to extract portions of the train data~\citep{nasr2023scalable}, etc.

\smallskip
We stress that, unlike other attacks against synthetic data~\cite{hayes2019logan, stadler2022synthetic, van2023membership}, we do not provide the adversary with any {\em side knowledge}.
That is, they have no access to the train/test data or even possession of data from the same distribution, no background information of the used generative approach, model, hyperparameters, nor model updates or gradients.
The adversary is also agnostic to the dataset type and the specific use case/downstream task.

\subsection{Membership and Attribute Inference Attacks}
\label{subsec:mia}

\descr{Membership Inference.}
The adversary aims to determine whether a target record, $r_t$, was part of the train data~\citep{shokri2017membership, hayes2019logan}.
Using \emph{DifferenceAttack}, outlined in Algorithm~\ref{alg:DifferenceAttack}, the adversary succeeds in doing so with just two API calls to the metrics: one using $D_{synth}^{n^{\prime}}\cup r_t$ (with the target) and one using $D_{synth}^{n^{\prime}}$ (without the target).
If the outputs are identical -- i.e., IMS is unchanged -- the target is not a member; otherwise, the attacker can confidently infer that it is.

\descr{Attribute Inference.}
We frame attribute inference~\citep{yeom2018privacy, stadler2022synthetic} through the lens of multiple membership inference attacks, similar to previous work~\citep{guo2023analyzing, berrada2023unlocking, hayes2024bounding}.
We assume the adversary has access to a partial target record from the train data, which we denote as $r_t$.
In other words, they know all but one attribute of the target record and aim to infer the missing attribute.
We adapt \emph{DifferenceAttack} for this task:
the adversary makes $k$ calls, using different possible values for the unknown attribute, to the privacy metrics ($D_{synth}^{n^{\prime}}\cup r_{t_1}$, $D_{synth}^{n^{\prime}}\cup r_{t_2}$, $\ldots$, $D_{synth}^{n^{\prime}}\cup r_{t_k}$).
The correct attribute value corresponds to the call with highest IMS.
In principle, the attack could be extended to $t$ unknown attributes, as in~\citep{oprisanu2022on}; however, the chance of multiple reconstructed records increases, potentially reducing precision.

\begin{algorithm}[t]
	\footnotesize
	\caption{\emph{DifferenceAttack}}
	\label{alg:DifferenceAttack}
	\begin{algorithmic}[1]
		\Require
		\Statex Trained Generative Model, $G_{\overline{\theta}}$
		\Statex Privacy Metrics, $M$ (specifically, IMS)
		\Statex Size of train data, $n_{train}$
		\Statex Target record, $r_t$
		\Procedure{DifferenceAttack}{$G_{\overline{\theta}}$, $M$, $n_{train}$, $r_t$}
			\State Generate $S \gets G_{\overline{\theta}}.\text{sample}(n_{train})$		\ccomment{new synthetic data}
			\State Query $\text{IMS} \gets M(S)$		\ccomment{metrics}
			\State Query $\text{IMS}_t \gets M(S \cup r_t)$		\ccomment{metrics}
			\If{$\text{IMS} == \text{IMS}_t$}
				\State \Return False		\ccomment{non-member}
			\Else{}
				\State \Return True		\ccomment{member}
			\EndIf
		\EndProcedure

	\end{algorithmic}
\end{algorithm}

\subsection{Reconstruction Attack}
\label{sec:attack}
Next, we present \emph{ReconSyn}, a reconstruction attack aimed at recovering the outliers in the train data.
An overview of the attack is shown in Figure~\ref{fig:reconsyn}.
Reconstruction is a more challenging and impactful adversarial task than membership/attribute inference, as it does not assume access to a target record, yet, the adversary can recover entire records from the train data.

\emph{ReconSyn} starts by locating the regions with outliers through \emph{OutliersLocator} and then launches two subattacks: 1) \emph{SampleAttack}, which passively generates and evaluates samples drawn from the generative model, and 2) \emph{SearchAttack}, which actively and strategically examines the history of records generated in the first phase.
\emph{ReconSyn}'s pseudocode is provided in Algorithm~\ref{alg:ReconSyn}.

\begin{algorithm}[t!]
	\footnotesize
	\caption{\emph{ReconSyn}}
	\label{alg:ReconSyn}
	\begin{algorithmic}[1]
		\Require
		\Statex Trained Generative Model, $G_{\overline{\theta}}$
		\Statex Privacy Metrics, $M$ (namely, IMS, DCR, NNDR)
		\Statex Size of train data, $n_{train}$
		\Statex Size of train outliers, $n_{out}$
		\Statex SampleAttack rounds, $r_{sma}$
		\Statex SearchAttack target distances, $d_{sra}$
		\Procedure{OutliersLocator}{$G_{\overline{\theta}}$, $n_{train}$, $n_{out}$}
			\State Generate $S \gets G_{\overline{\theta}}.\text{sample}(3 \cdot n_{train})$		\ccomment{new synthetic data}
			\State Init., fit, and predict $c_{out} \gets GM.\text{fit\_predict}(S)$		\ccomment{Gaussian Mixture}
			\State Select $c_{out} \gets \max \{c \subseteq c_{out} : \sum_{c' \in c} |c'| \leq n_{out}\}$
			\Statex 	\ccomment{outliers clusters}

			\State \Return $c_{out}, GM$
		\EndProcedure

		\Procedure{SampleAttack}{$G_{\overline{\theta}}$, $M$, $GM$, $r_{sma}$, $n_{train}$, $c_{out}$}
			\State Init. $R_{sma} \gets \emptyset$		\ccomment{SMA recon. outliers to the empty set}
		 	\State Init. $H_{out} \gets \emptyset$		\ccomment{history to the empty set}
			\For{$r$ in $r_{sma}$}		\ccomment{iterate over number of rounds}
				\State Generate $S \gets G_{\overline{\theta}}.\text{sample}(n_{train})$		\ccomment{new synthetic data}
				\State Select $S \gets \{S[i] \;|\; GM.\text{predict}(S)[i] \in c_{out}\}$
				\Statex	\ccomment{outliers candidates}
				\State Filter $S \gets S \; \backslash \; H_{out}$		\ccomment{candidates out from history}
				\State Query $\text{dists} \gets M(S)$		\ccomment{metrics (augment if necessary)}
				\State Update $R_{sma} \gets R_{sma} \; \cup \; \{S[i] \;|\; dists[i] = 0 \}$
				\Statex		\ccomment{recon. outliers}
				\State Update $H_{out} \gets H_{out} \; \cup \; \{\text{Zip}(S, dists)\}$		\ccomment{history}
			\EndFor

			\State \Return $R_{sma}, H_{out}$
		\EndProcedure

		\Procedure{SearchAttack}{$M$, $GM$, $d_{sra}$, $c_{out}$, $H_{out}$}
		 	\State Init. $R_{sra} \gets \emptyset$		\ccomment{SRA recon. outliers to the empty set}
			\State Select and sort $H'_{out} \gets \{H_{out}[i] \;|\; dists[i] \le d_{sra}\}$
			\Statex		\ccomment{trim history}
			\For{$s, dist_s$ in $H'_{out}$}		\ccomment{iterate over history}
				\State Build $N_s \gets \{ s \; \text{with} \; s[i] \; \text{modified}, \; \forall i \in [1, \text{length}(s)] \}$
				\Statex	\ccomment{record neighboring dataset}
				\State Filter $N_s \gets N_s \; \backslash \; H_{out}$		\ccomment{candidates out from history}
				\State Query $\text{dists} \gets M(N_s)$		\ccomment{metrics (augment if necessary)}
				\State Update $H_{out} \gets H_{out} \; \cup \; \{\text{Zip}(N_s, dists)\}$		\ccomment{history}
				\State Select $c_s \gets \{i \;|\; dists[i] \le dist_s\}$		\ccomment{find cols yet to be recon.}
				\For{$c_{s_i}$ in $c_s$}		\ccomment{iterate over cols yet to be recon.}
					\State Build $CC_{s} \gets \{s_{i} \;|\; \forall \, val \in \text{Support}(c_{s_i})\}$
					\Statex	\ccomment{col closer candidates}
					\State Select $CC_{s} \gets \{CC_{s}[j] \;|\; GM.\text{predict}(CC_{s})[j] \in c_{out}\}$
					\Statex	\ccomment{outliers candidates}
					\State Filter $CC_{s} \gets CC_{s} \; \backslash \; H_{out}$		\ccomment{candidates out from history}
					\State Query $\text{dists} \gets M(CC_{s})$		\ccomment{metrics (augment if necessary)}
					\State Update $R_{sra} \gets R_{sra} \; \cup \; \{CC_{s}[j] \;|\; dists[j] = 0 \}$
					\Statex	\ccomment{recon. outliers}
					\State Update $H_{out} \gets H_{out} \; \cup \; \{\text{Zip}(CC_{s}, dists)\}$		\ccomment{history}
				\EndFor
			\EndFor

			\State \Return $R_{sra}$
		\EndProcedure

	\end{algorithmic}
\end{algorithm}

\descr{\em OutliersLocator.}
As mentioned, the adversary uses \emph{OutliersLocator} to identify regions with underrepresented records or outliers.
This involves generating a large synthetic data sample, fitting a Gaussian Mixture model, and selecting the smallest isolated clusters.
We implement two strategies to select outliers and cover a wide set of scenarios as they could lie outside or within the cluster(s).

\descr{\em SampleAttack.}
In each round, we generate synthetic data, identify potential outliers using \emph{OutliersLocator}, and remove records examined in previous rounds. %
The attack then queries the metrics API for exact matches (if all tests pass) and adds the queried data to the history.

Following the second counter-example in Section~\ref{sec:examples}, the aggregate scores can be tricked to expose exact distances.
Specifically, we can determine the distance between a target synthetic record and the nearest train data counterpart by conditionally generating or augmenting the input synthetic data.
One effective way is to submit the target point alongside, e.g., 100 copies of a frequently appearing record whose closest distance we know.
This essentially simulates a scenario where the metrics API reveals individual distances to all submitted data (which we %
do in lines 14, 26, and 33), allowing the detection of matches with 100\% confidence.

\descr{\em SearchAttack.}
Informally, the intuition is to select close records from the history and `shake' or `fix' them one column at a time until an exact match is found.
For a specific record, first, we identify columns that have not yet been reconstructed using its neighboring dataset, which is a square matrix where each row differs in a single column value.
Then, we iteratively test possible values for these columns, filtering out records through \emph{OutliersLocator} and the history.
Ultimately, this leads to another match.

We expand on the definition of the record's neighboring dataset and its role in identifying the unreconstructed columns for that record (lines 24--28).
The neighboring dataset could be built by creating a square matrix by duplicating the record $d$ times ($d$ is the number of columns), and then altering the values along the diagonal by some amount.
When this dataset is fed into the metrics API, the distances to the columns that have been accurately reconstructed will most likely increase since their values have been moved away from an exact match.
Consequently, this indicates that the unchanged/closer columns must be corrected.

Finally, we look closer at determining the correct values for these columns (lines 29--35).
To reduce the number of calls to the metrics API, which could be significant if all possible combinations were enumerated, we use a greedy strategy.
This involves iterating over the columns one at a time and building all potentially closer candidates by going over the possible values for the current column.
Additionally, to further minimize the number of API calls, we use \emph{OutliersLocator} and recorded history to filter out unsuitable candidates.

\descr{Efficiency.}
One could argue that %
using the Hamming distance limits the adversary's efficiency, as it does not provide any sense of direction (at all times, any value is either an exact match or not).
Nonetheless, our attack achieves strong performance in reconstructing the train outliers (as shown in Section~\ref{sec:rec}) and is computationally practical.
In all our experimental settings, every phase runs in less than 24 hours on an m4.4xlarge (16 CPUs, 64GB RAM) AWS instance.

The attack includes querying a single trained generative model up to 5,000 times.
This is well within the scope of similar privacy attacks, which typically involve training (which takes significantly longer than querying a trained model) anywhere from 1,000~\cite{stadler2022synthetic, annamalai2023linear} and 10,000~\citep{tramer2022debugging} to 1,000,000~\citep{nasr2021adversary} models.

\descr{Why Reconstruction and Why Outliers?}
Besides membership and attribute inference, we focus on the more ambitious reconstruction task, as, arguably, this is significantly more powerful and consequential.
\emph{ReconSyn} exposes {\em all} (sensitive) attributes, unequivocally demonstrating that similarity-based privacy guarantees are not fit for purpose.
In fact, even if the attack successfully reconstructed only a few train outliers with high precision, it would still constitute a severe privacy violation for the affected individuals~\citep{carlini2022membership}.

Reconstruction implies the ability to single individuals out and enable their identification or link them to the real data.
This, in turn, means that the process of generating synthetic data and guaranteeing its privacy has failed at least two of the three privacy guarantees outlined by EU's Article 29 Data Protection Working Party~\citep{eu2014opinion}, namely, singling out and linkability.
Thus, the process cannot be considered anonymous as per the GDPR.

We target the underrepresented regions in the train data as they might correspond to the most vulnerable individuals.
In fact, regulators like the UK's ICO~\citep{ico2022privacy} have explicitly highlighted the importance of protecting outliers.
Moreover, they are inherently more difficult to model accurately, which makes their reconstruction more challenging.
We elaborate on this further in Appendix~\ref{app:all}.

\begin{table}[t!]
	\vspace{-0.3cm}
	\small
	\centering
	\setlength{\tabcolsep}{4pt}
	\begin{tabular}{l|rrr|r}
		\toprule
			\textbf{Model}				& %
														\emph{Groundhog} 		& \emph{Querybased} 	&	\emph{DOMIAS}		& \emph{Difference}	\\
														& \cite{stadler2022synthetic}	& \citep{houssiau2022tapas}		& \citep{van2023membership}	&	\emph{Attack}		\\
		\midrule
			\textbf{PrivBayes}		&	0.51							& 0.57								& 0.55						& \textbf{1.00}			\\
			\textbf{MST}					&	0.55							& 0.54								& 0.51						& \textbf{1.00}			\\
			\textbf{DPGAN}				&	0.95							& 0.95								& 0.66						& \textbf{1.00}			\\
			\textbf{PATE-GAN}			&	0.57							& 0.55								& 0.55						& \textbf{1.00}			\\
			\textbf{CTGAN}				&	0.65							& 0.80								& 0.56						& \textbf{1.00}			\\
	 \midrule
	 		\textbf{\#Calls}			&	0.5k--2k					& 0.5k--2k						& 0.5k--2k				& \textbf{2}				\\
			\textbf{Runtime}			&	2h--121h					& 2h--124h						& 1h-56h					& \textbf{$<$1s}		\\
	 \bottomrule
	\end{tabular}
	\vspace{2pt}
	\caption{Comparison between \emph{DifferenceAttack} and state-of-the-art membership inference attacks on a target train outlier, \emph{Adult Small}: AUC score, number of API calls, and runtime.}
	\label{tab:mias}
	\vspace{-0.2cm}
\end{table}

\begin{table*}[t!]
	\vspace{-0.3cm}
	\small
	\centering
	\begin{tabular}{l|r|rr|rr|rr|rr}
		\toprule
			\textbf{Model}			& {\em 2d Gauss} 	& \multicolumn{2}{c|}{\em Adult Small}	& \multicolumn{2}{c|}{\em Adult}	& \multicolumn{2}{c|}{\em Census}	& \multicolumn{2}{c}{\em MNIST} 		\\
													& \emph{Sample}		& 1 call 				& \emph{Sample}					& \emph{Sample}	& \emph{Search}		& \emph{Sample} & \emph{Search}		& \emph{Sample}			& \emph{Search}	\\
		\midrule
			\textbf{Oracle}			& 0.95						& 							&												&								&									&								&									&										&								\\
			\textbf{PrivBayes}	&									& 0.10					& 1.00									& 0.44					& 0.95						& 0.54					&	0.98						& 0.00							& 0.99					\\
			\textbf{MST}				&									& 0.17					& 1.00									& 0.05					& 0.90						&	0.84					&	0.99						& 0.00							& 0.97					\\
			\textbf{DPGAN}			&									& 0.11					& 0.96									& 0.02					& 0.78						&	0.15					&	0.82						& 0.00							& 0.97					\\
			\textbf{PATE-GAN}		&									& 0.08					& 1.00									& 0.02					& 0.81						&	0.37					&	0.83						& 0.00							& 0.97					\\
			\textbf{CTGAN}			&									& 0.10					& 0.99									& 0.00					& 0.80						&	0.74					&	0.90						& 0.00							& 0.80					\\
	 \bottomrule
	\end{tabular}
	\vspace{2pt}
	\caption{Overview of \emph{ReconSyn}'s reconstruction success rate (\emph{SampleAttack} and \emph{SearchAttack}) with different datasets and against different models (all DP models are trained with $\epsilon=\infty$). {\em NB:} `1 call' refers to {\em SampleAttack} with only one API call to the metrics.}
	\label{tab:summary}
	\vspace{-0.2cm}
\end{table*}

\section{Experimental Evaluation}
\label{sec:rec}
In this section, we evaluate the effectiveness of our privacy attacks.
First, we show \emph{DifferenceAttack}'s perfect performance in membership and attribute inference and compare it to existing methods.
We then demonstrate that \emph{ReconSyn} successfully recovers the train outliers in different settings.
More specifically, we measure the performance of \emph{ReconSyn} (\emph{SampleAttack} and \emph{SearchAttack}) against PrivBayes, MST, DPGAN, PATE-GAN, and CTGAN (in non-private settings, i.e., $\epsilon=\infty$) on increasingly more complex datasets (\emph{2d Gauss}, \emph{Adult Small}, \emph{Adult}, \emph{Census}, \emph{MNIST}, described in Appendix~\ref{app:data}).

\subsection{\emph{DifferenceAttack}}
\label{subsec:mem}
We start by demonstrating that the simple \emph{DifferenceAttack} achieves better results compared to state-of-the-art attacks (SOTA) with considerably less computation (see Table~\ref{tab:mias}) owing to the leakage coming from the privacy metrics (and their stronger implicit threat model).

\subsubsection{Membership Inference}
\emph{DifferenceAttack} achieves perfect results vs. all generative models and takes under a second (see Table~\ref{tab:mias}) by making just two calls to the privacy metrics, as discussed in Section~\ref{subsec:mia}.
Apart from the target record $r_t$ (the complete record, including all attributes), the adversary does not have access to any side information. %

By comparison, SOTA black-box membership inference attacks~\cite{stadler2022synthetic,houssiau2022tapas,van2023membership}
typically assume access to $r_t$, representative data, and the model's training algorithm.
The adversary fits several {\em shadow} models -- trained to mimic the model's behavior -- using datasets that either include or exclude $r_t$ to assess its influence.
Then, they extract features from synthetic datasets generated from these `in' and `out' shadow models, with a subset to train a classifier and the rest to determine if $r_t$ was part of the original train data.
In Groundhog~\citep{stadler2022synthetic}, the adversary featurizes synthetic datasets by extracting min, max, mean, mode, and variance for each column (or most rare/common categories), while Querybased~\citep{houssiau2022tapas} involves running a collection of random queries.
DOMIAS~\citep{van2023membership} directly extracts the inclusion probability by comparing the densities of synthetic data and reference data not used in the generative model training.

To estimate the AUC from GroundHog~\cite{stadler2022synthetic}, Querybased~\cite{houssiau2022tapas}, and DOMIAS~\cite{van2023membership}, we use 1,200 models (for Groundhog and Querybased) to train the classifier and 800 models (for all three attacks) for testing when attacking PrivBayes and MST, and 300 for training/200 for testing against the GANs, as they take much longer to converge.
This is comparable to Groundhog's default 1,000 and Querybased's 3,500 models, even though we test more numerous and more complex generative model implementations. %
(We do not expect the AUC scores to significantly change if we used the default number of models, as results mostly plateau.)
As shown in the first three columns in Table~\ref{tab:mias}, with a couple of exceptions (Groundhog vs. DPGAN and Querybased vs. DPGAN and CTGAN), the three attacks achieve AUC results below 0.66 and can take over 120 hours to run, particularly against the GANs.

\subsubsection{Attribute Inference}
As discussed in Section~\ref{subsec:mia}, \emph{DifferenceAttack} can perfectly (AUC score = 1.0) predict the unknown attribute of a partial train record with $k$ calls to the metrics, where $k$ is the number of distinct categories of the unknown attribute.
This process takes just a few seconds. %

By contrast, SOTA attribute inference attacks do not achieve very high AUC scores, being reported as at most 0.75 on PrivBayes~\citep{annamalai2023linear} and 0.5 for MST~\citep{houssiau2022tapas} and CTGAN~\citep{houssiau2022tapas, annamalai2023linear} ($\epsilon = \infty$ for PrivBayes/MST).
Thus, we do not see the benefit of reproducing experiments with the SOTA attribute inference attacks.
Similarly, we do not adapt the three membership inference attacks to attribute inference via multi-membership inference would require hundreds of hours and is unlikely to yield stronger results anyway.

\subsection{\emph{SampleAttack}}
Next, we move on to reconstruction and evaluate the performance of the first \emph{ReconSyn} subattack, i.e., \emph{SampleAttack}.

\subsubsection{\emph{SampleAttack} with One Call} %
We start with \emph{SampleAttack} on all datasets/models, limiting the number of metrics API calls to just one to simulate a strict scenario where only a single synthetic dataset is shared with its accompanying metrics.
Specifically, we generate one synthetic dataset and use it for both fitting \emph{OutliersLocator} to select potential outliers as well as making an API call to the metrics.
Since we cannot make further calls, we predict records as outliers based on the number of duplicates among the selected records using the IMS score as an upper limit, reasoning that if a record is generated multiple times, it is more likely to be in the train data~\citep{dick2023confidence}.

We report the results for \emph{Adult Small} in 1 call column of Table~\ref{tab:summary}.
In this setting, one call is enough to reconstruct 8\% to 17\% of outliers with precision ranging from 13\% and 32\%.
While reconstructing any outliers could be considered serious privacy leakage, the adversary's confidence is quite low.
Moreover, a larger number of calls are needed for all other datasets as we fail to recover any outliers.
To ease presentation, we do not report these results in Table~\ref{tab:summary}.

\subsubsection{\emph{SampleAttack} with Multiple Calls}
We then evaluate the full \emph{SampleAttack} subattack, running 1,000 rounds on all datasets except for \emph{MNIST}, where we use 5,000.
Overall, we obtain mixed results: regardless of the target model, the attack is very successful on \emph{2d Gauss} and \emph{Adult Small}, reconstructing at least 95\% of train outliers.
Whereas, as we discuss below, the subattack struggles for the other three datasets (also see Table~\ref{tab:summary}).

\descr{\emph{2d Gauss}.}
For \emph{2d Gauss}, we attack the `oracle' (introduced in CE3 in Section~\ref{sec:examples}), which possesses knowledge of the generative process.
Even though no generative model has been exposed to train data, and the oracle has no memory of the synthetic data it generated, \emph{SampleAttack} reconstructs 95\% of train outliers due to leakage from the privacy metrics.

\begin{figure}[t!]
	\vspace{-0.3cm}
	\centering
	\includegraphics[width=0.99\linewidth]{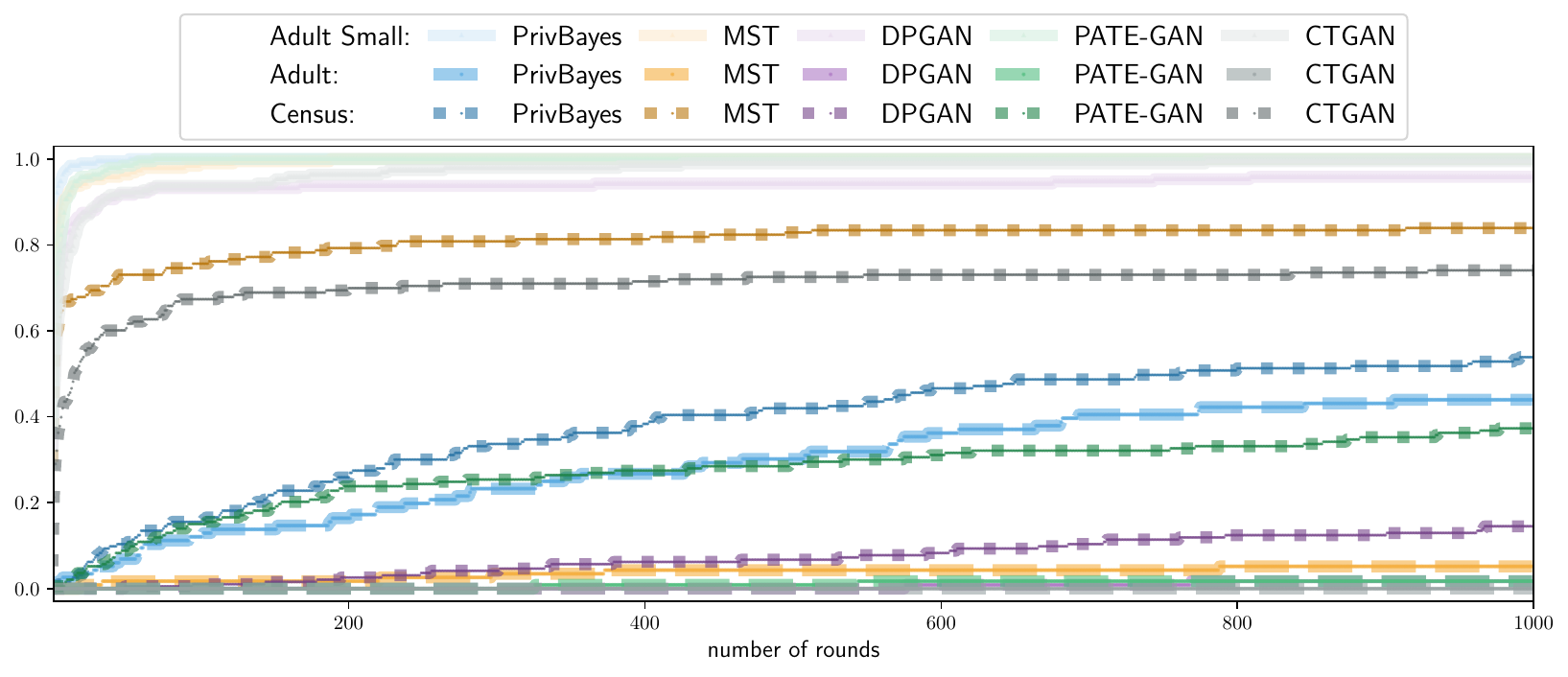}
	\caption{Reconstruction success rate of train outliers for increasing \emph{SampleAttack} rounds, \emph{Adult Small}, \emph{Adult}, and \emph{Census}.}
	\label{fig:adult_n_recon}
\end{figure}

\descr{\emph{Adult Small}.}
For \emph{Adult Small}, we use \emph{SampleAttack} against all five generative models and report the number of reconstructed outliers %
in Figure~\ref{fig:adult_n_recon} (top five lines).
For PrivBayes, MST, and PATE-GAN, the attack quickly reconstructs about 90\% outliers after just ten rounds and eventually reaches 100\%.
For DPGAN and CTGAN, the attack plateaus at around 85\% after 40 rounds, but by round 1,000, it improves and achieves 96\% and 99\%, respectively.
We believe that \emph{SampleAttack} is extremely successful on this dataset because of its small cardinality ($10^5$).

\descr{\emph{Adult}.}
The models are much less likely to memorize and reproduce individual data points on \emph{Adult}, which has twice the number of columns and cardinality of $10^{15}$.
Indeed, apart from PrivBayes, \emph{SampleAttack} only recovers 5\% of the outliers -- see Table~\ref{tab:summary} and Figure~\ref{fig:adult_n_recon} (bottom five lines).

\descr{\emph{Census}.}
Even though \emph{Census} has roughly twice the columns/rows and much higher cardinality, \emph{SampleAttack} is more successful (excluding PrivBayes on \emph{Adult}), recovering on average 53\% outliers (see middle lines in Figure~\ref{fig:adult_n_recon}).
Interestingly, attacking CTGAN yields better results than PrivBayes.
Also, the recovery rate follows a linear trend (vs.~logarithmic for \emph{Adult Small}).

\begin{figure}[t!]
	\centering
	\includegraphics[width=0.99\linewidth]{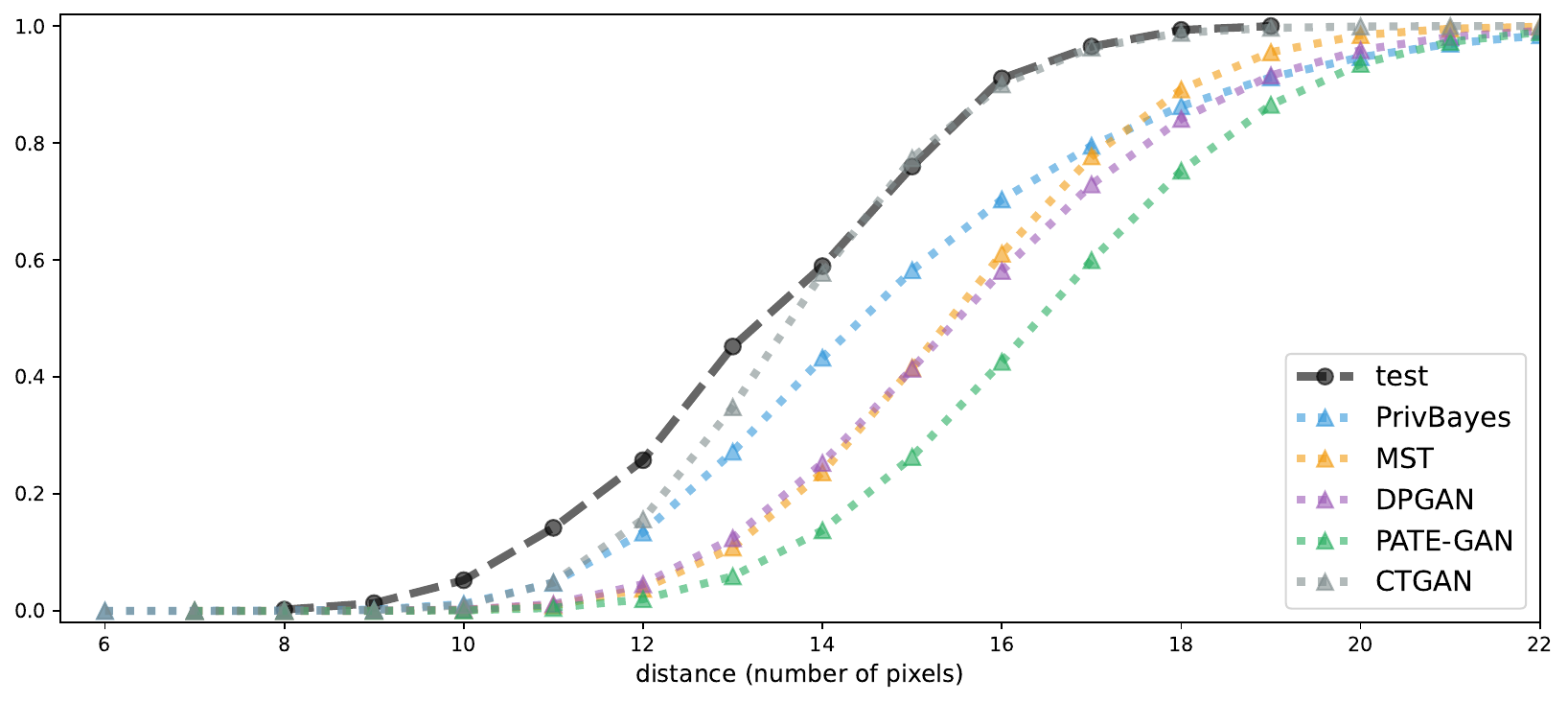}
	\caption{CDF of similarity distances between outliers in train-test and train-synthetic datasets observed by \emph{SampleAttack}, \emph{MNIST}.}
	\label{fig:mnist_cdf}
	\vspace{-0.2cm}
\end{figure}

\descr{\emph{MNIST}.}
Finally, looking at \emph{MNIST}, which has even higher dimensionality/cardinality, \emph{SampleAttack} fails completely and does not reconstruct even a single outlier.
In fact, in Figure~\ref{fig:mnist_cdf}, we see that CTGAN and PrivBayes generate images with the highest similarity to the real ones but still at a Hamming distance of at least 6 (i.e., number of different pixels).
All models, however, create outliers further away from the train data compared to the distances between test and train.
This confirms that all privacy tests pass and that test/synthetic datasets do not contain train data copies.

\subsection{\emph{SearchAttack}}
Finally, we run the follow-up \emph{SearchAttack} subattack on all models where \emph{SampleAttack} achieves less than 95\% reconstruction success, i.e., we attack all five models on \emph{Adult}, \emph{Census}, and \emph{MNIST}.
We run it for up to $1/4$ of the distances; in other words, we go through the history and try to `fix' at most four columns of \emph{Adult}/\emph{Census} and 16 of \emph{MNIST}.
While widening the search to further distances could lead to better results, we limit the number of computations to keep our attack efficient.
In all cases, we manage to reconstruct over 78\% of all train outliers; see Table~\ref{tab:summary}.

\descr{\emph{Adult}.}
For \emph{Adult}, \emph{SearchAttack} easily recovers the majority of train outliers, between 78\%--95\%.
The attack is both more effective and efficient on the graphical models since it does not need to go far in history -- only a couple of distances (or columns).
Most likely, this is due to: 1) \emph{SampleAttack} already being more successful for these two models, generating more diverse data history, and 2) graphical models tending to overperform GANs on low-dimensional datasets like \emph{Adult} and simple downstream tasks like marginal preservation~\citep{tao2022benchmarking, ganev2024graphical}.
Nonetheless, \emph{SampleAttack} reconstructs most outliers against the GAN models, too, even though it requires searching further back (distance of 4).

\begin{figure}[t!]
	\vspace{-0.3cm}
	\centering
	\begin{subfigure}{0.627\linewidth}
		\includegraphics{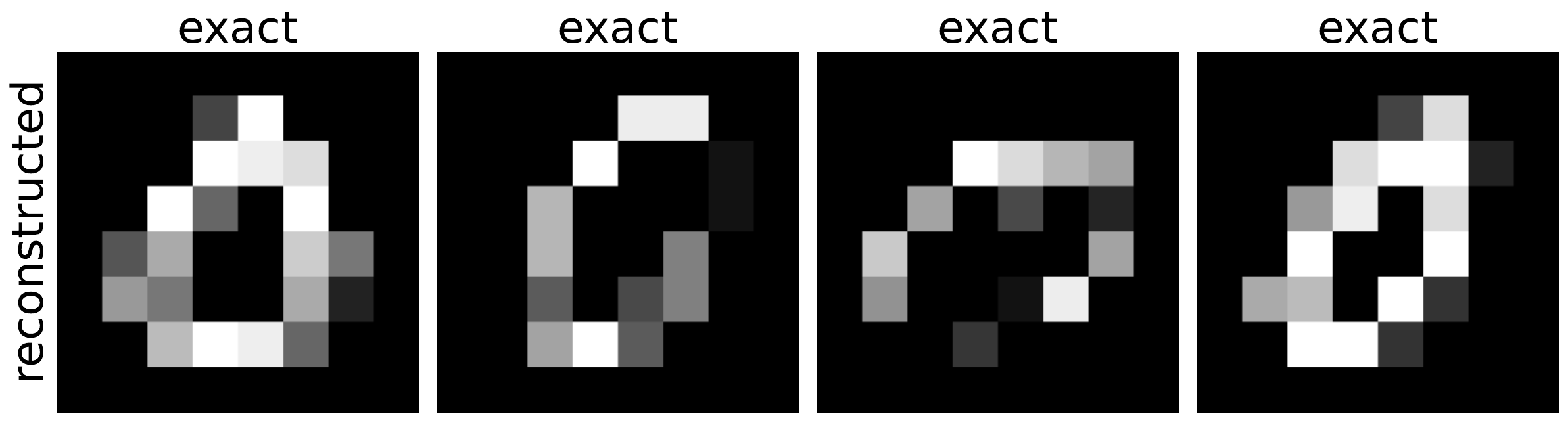}
		 \caption{PrivBayes, exact}
	 \end{subfigure}
	 \begin{subfigure}{0.313\linewidth}
		 \includegraphics{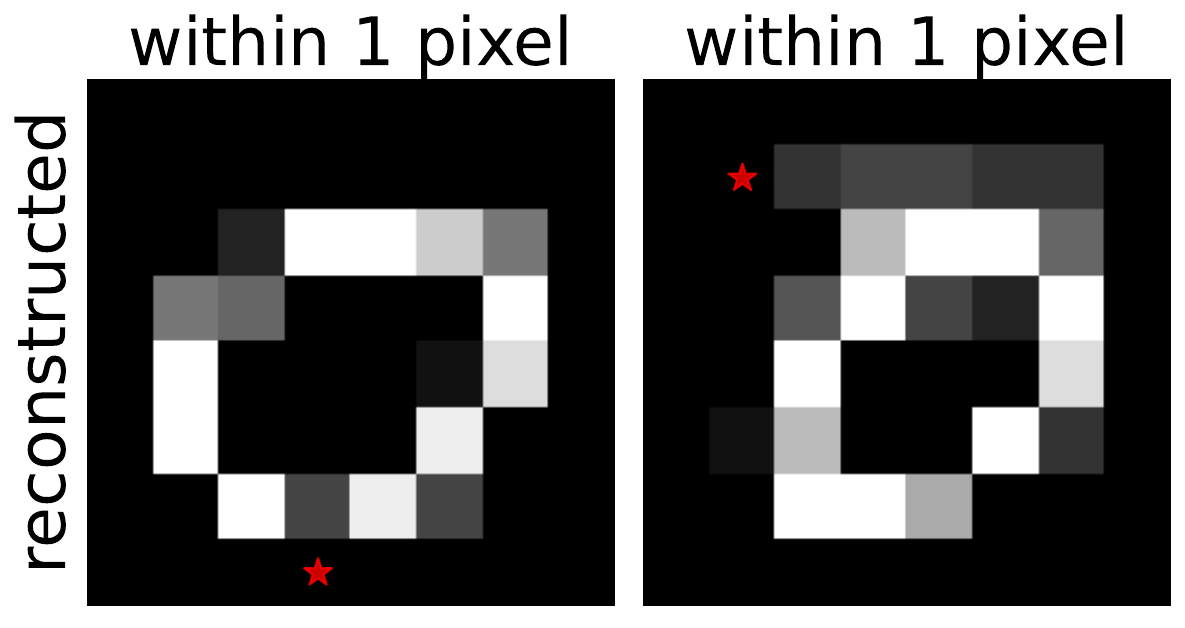}
		 \caption{PrivBayes, 1 pixel}
	 \end{subfigure}\\[0.75ex]
	 \begin{subfigure}{0.627\linewidth}
		 \includegraphics{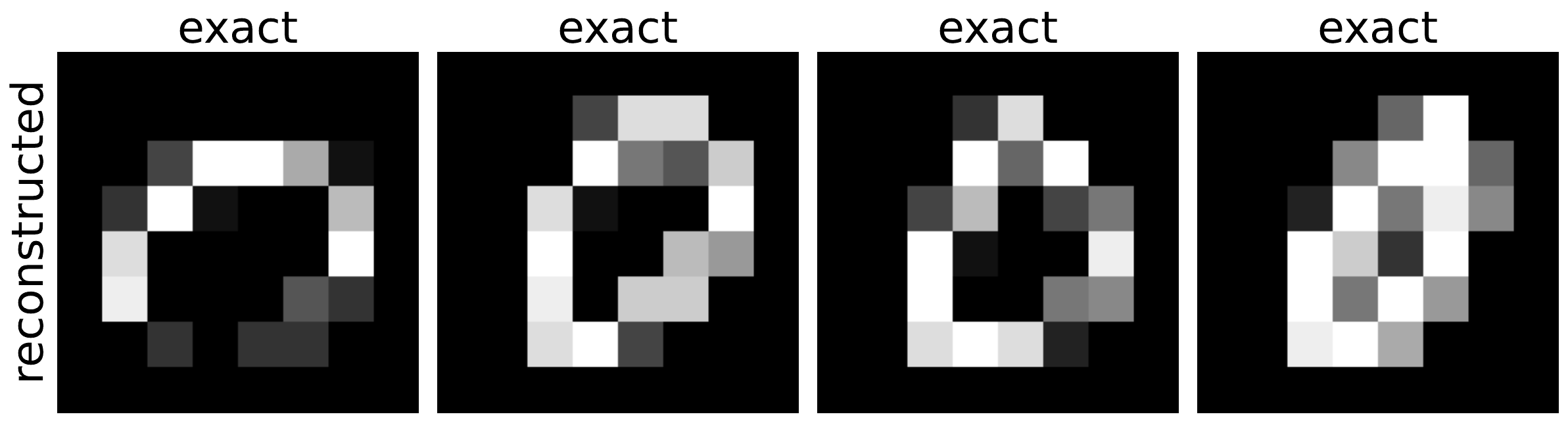}
		 \caption{MST, exact}
	 \end{subfigure}
	 \begin{subfigure}{0.313\linewidth}
		 \includegraphics{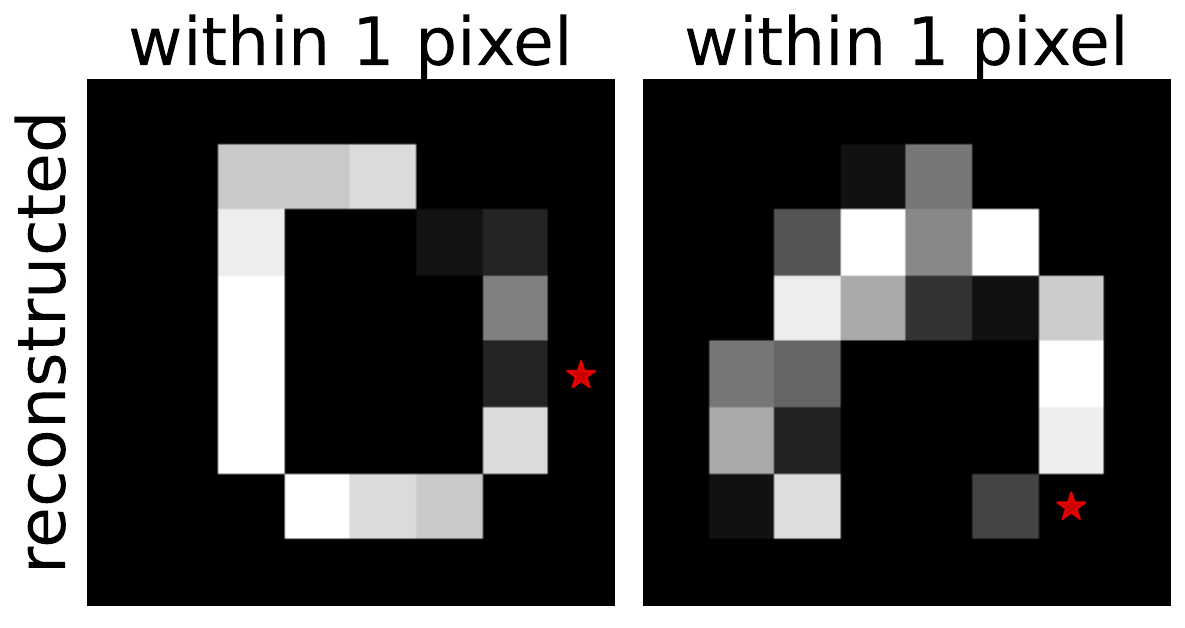}
		 \caption{MST, 1 pixel}
	 \end{subfigure}\\[0.75ex]
	 \begin{subfigure}{0.627\linewidth}
		 \includegraphics{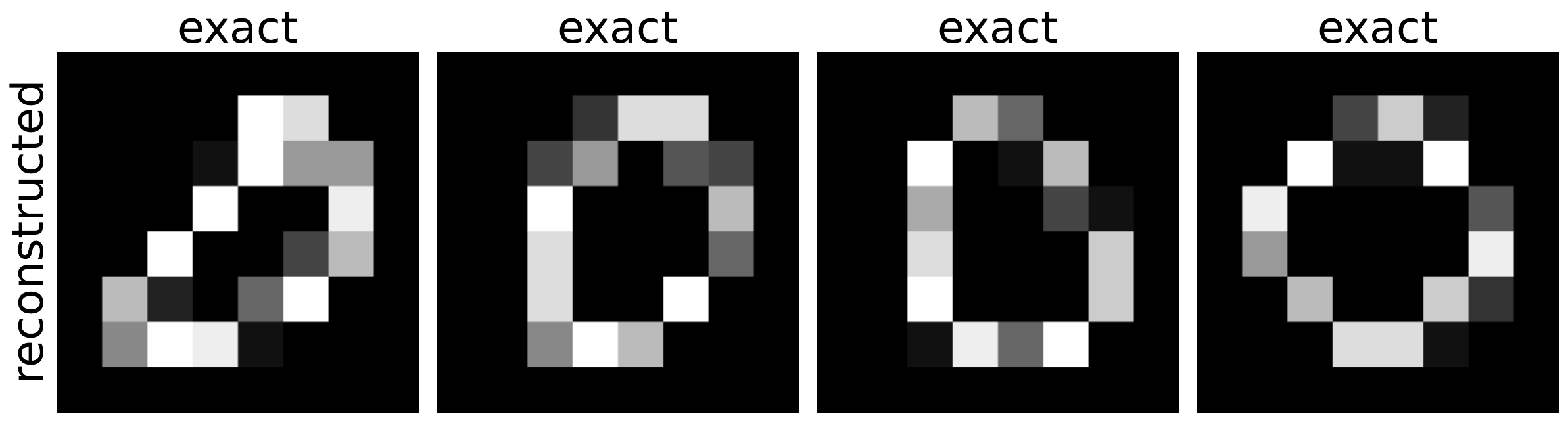}
		 \caption{DPGAN, exact}
	 \end{subfigure}
	 \begin{subfigure}{0.313\linewidth}
		 \includegraphics{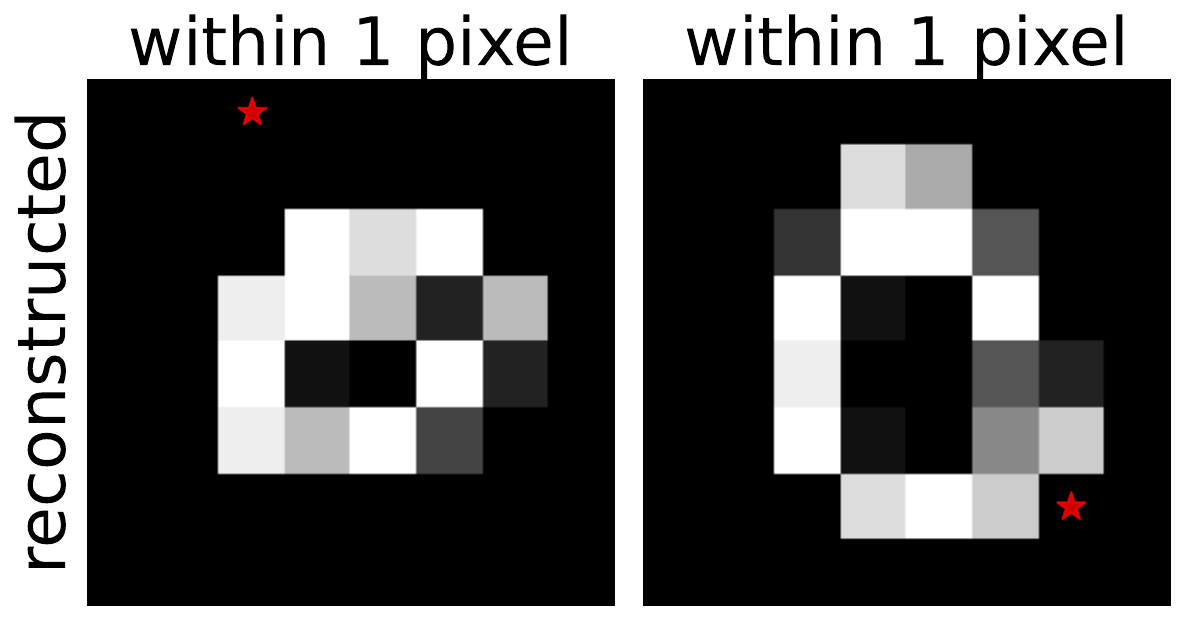}
		 \caption{DPGAN, 1 pixel}
	 \end{subfigure}\\[0.75ex]
	 \begin{subfigure}{0.627\linewidth}
		 \includegraphics{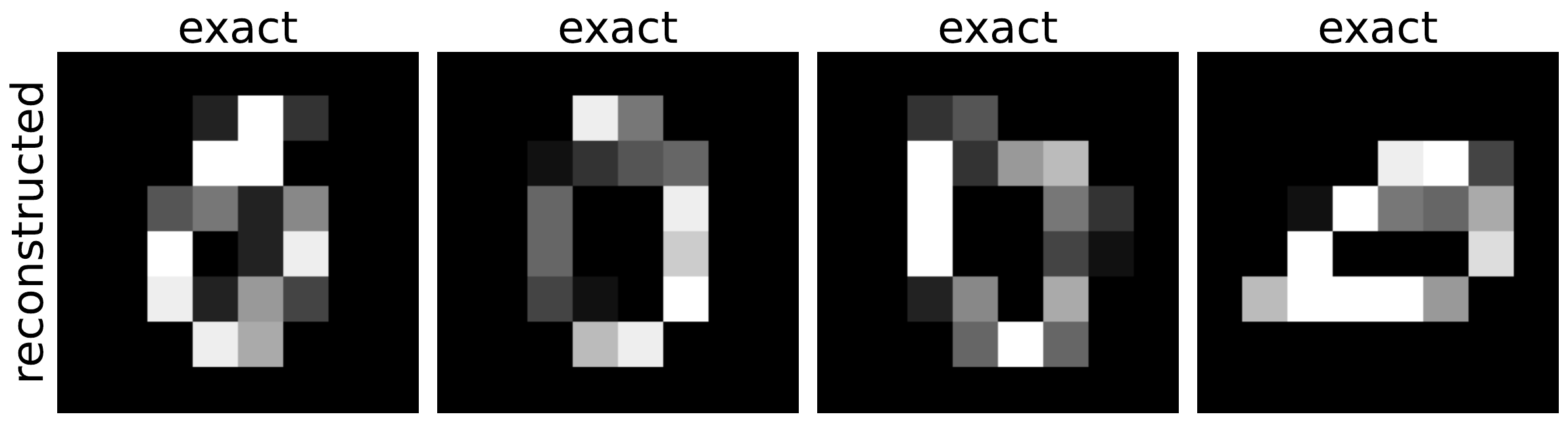}
		 \caption{PATE-GAN, exact}
	 \end{subfigure}
	 \begin{subfigure}{0.313\linewidth}
		 \includegraphics{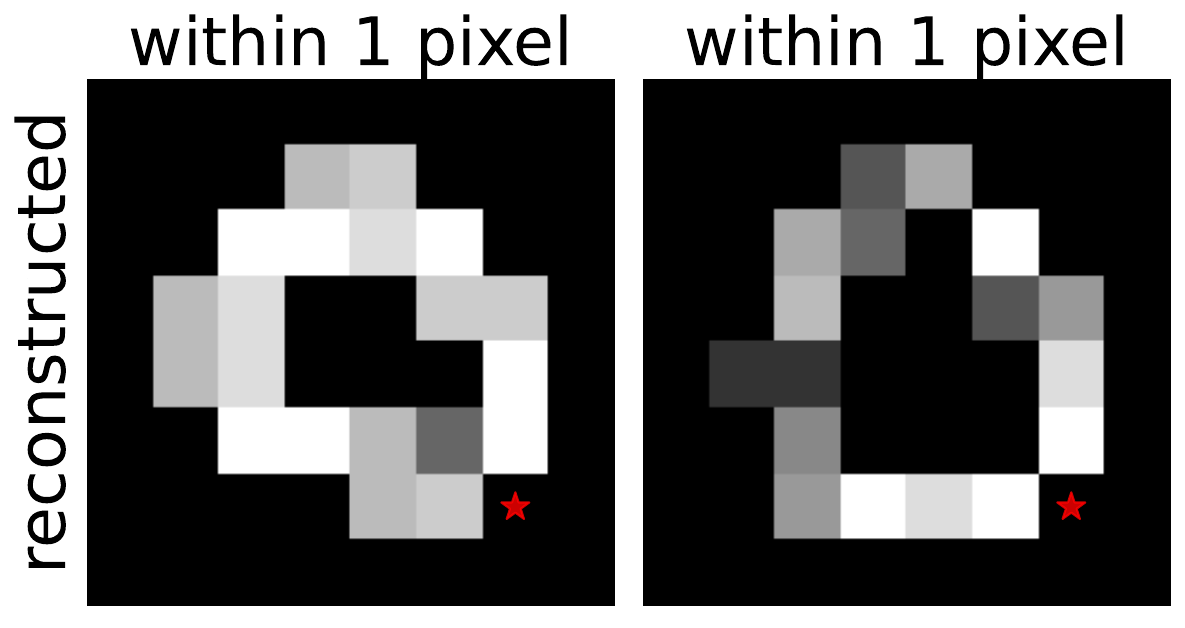}
		 \caption{PATE-GAN, 1 pixel}
	 \end{subfigure}\\[0.75ex]
	 \begin{subfigure}{0.627\linewidth}
		 \includegraphics{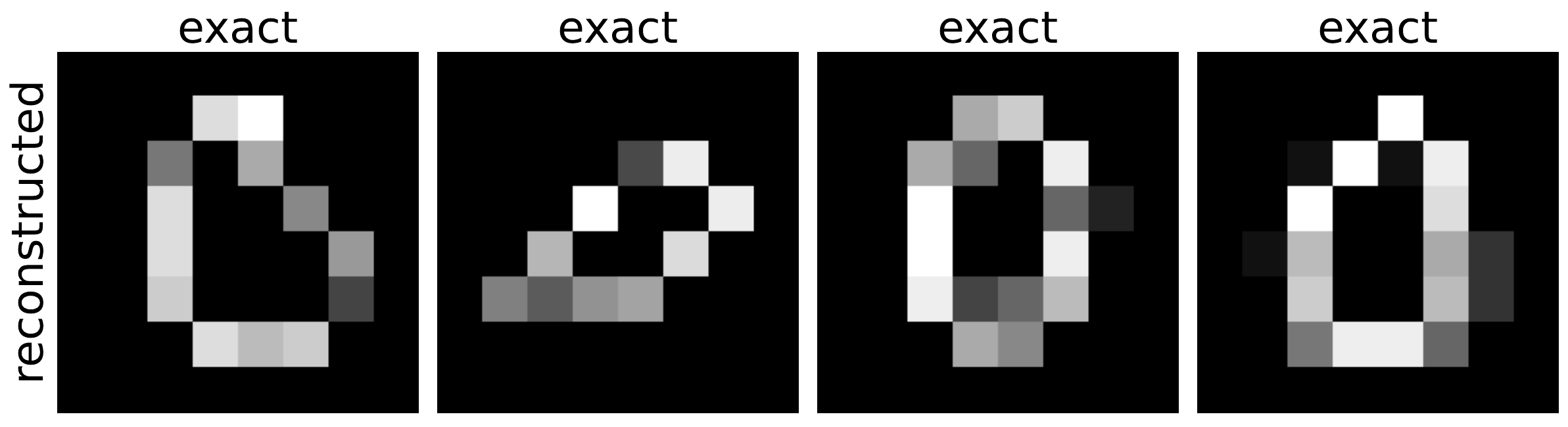}
		 \caption{CTGAN, exact}
	 \end{subfigure}
	 \begin{subfigure}{0.313\linewidth}
		 \includegraphics{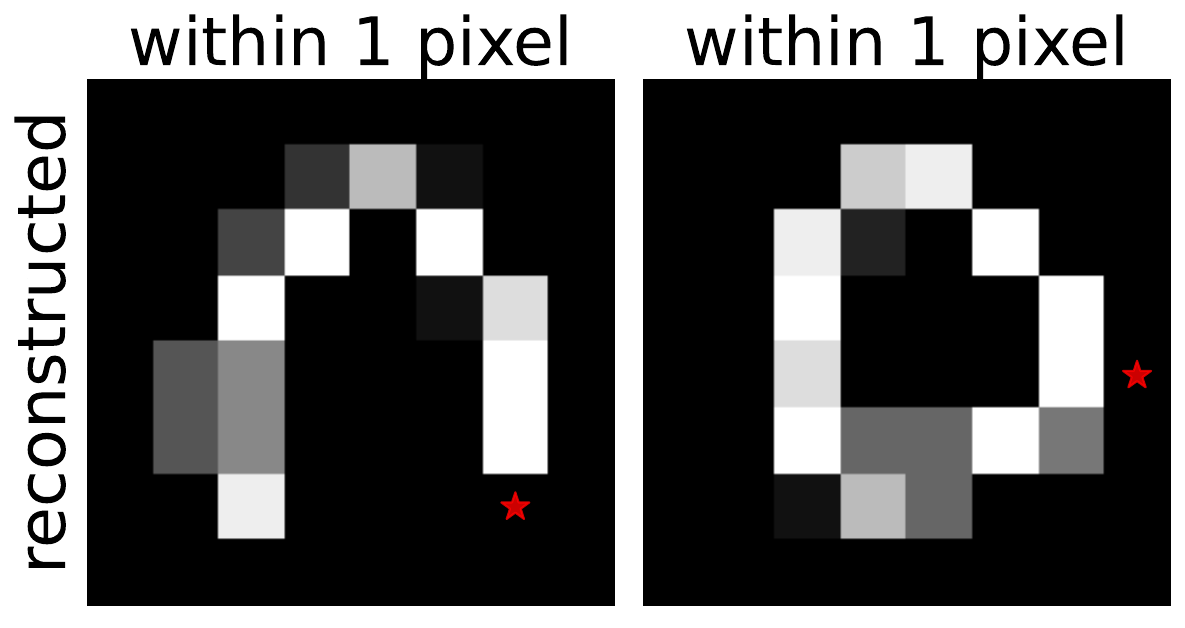}
		 \caption{CTGAN, 1 pixel}
	 \end{subfigure}\\[1ex]
	 \caption{Reconstructed train outliers by \emph{SearchAttack}, \emph{MNIST}.}
	 \label{fig:mnist_train_recon}
	\vspace{-0.2cm}
\end{figure}

\descr{\emph{Census}.}
\emph{SearchAttack}'s performance on \emph{Census} is similar -- it reconstructs more outliers vs.~the graphical models (99\%) than the GANs (85\% on average).
Even though attacking PrivBayes starts at a disadvantage compared to CTGAN, \emph{SearchAttack} manages to recover more outliers.
Again, this could be because PrivBayes generates richer history or potential overfitting of CTGAN.

\descr{\emph{MNIST}.}
Despite {\em MNIST}'s high cardinality, \emph{SearchAttack} reconstructs more than 80\% of the train outliers.
To reduce the search space, the adversary can be strategic, e.g., excluding some pixels (e.g., the ones on the sides of the image or the `frame') by setting their value to 0 after observing the common pattern and generating a collection of potential outliers.
This restricts the adversary's capability, and they cannot fully reconstruct 21 out of the 488 outliers.
Nonetheless, this could be considered a good tradeoff vis-\`a-vis the number of saved computations, a factor of $480 = 30 \cdot 16$ (30 fixed pixels, 16 combinations per bin) per search.
In the last column in Table~\ref{tab:summary}, we report the number of train outliers reconstructed exactly and those within 1 pixel (even though the adversary can easily get an exact train data match by running the attack for one step without any restrictions).
A subset of the recovered digits is shown in Figure~\ref{fig:mnist_train_recon}.

Overall, \emph{SearchAttack} is very successful, despite \emph{SampleAttack}'s failure to recover any outliers.
Attacking all models, except for CTGAN, results in reconstructing at least 97\% of outliers.
We believe this high score is due to the generators' ability to create diverse images not too dissimilar from the outliers (as shown in Figure~\ref{fig:mnist_cdf}).
Conversely, although CTGAN generates the closest images, that does not recover more outliers.
This might be due to mode collapse or its embedding strategy for categorical columns (both DPGAN and PATE-GAN use simple one-hot encoding).
Unsurprisingly, out of the restricted 21 outliers, attacking CTGAN leads to recovering only six compared to at least 16 for the other models.

\subsection{Take-Aways}
We show that \emph{DifferenceAttack} achieves perfect accuracy for membership and attribute inference with only a handful of API calls to the metrics. %
We also show that \emph{ReconSyn} successfully reconstructs at least 78\% of the train outliers with all tested models and datasets.
The \emph{SearchAttack} subattack performs better on lower-dimensional datasets but fails to recover any records for \emph{MNIST}, while \emph{SearchAttack} achieves an average of 90\% success on the wider datasets and is slightly more successful against graphical models.

\section{Related Work}
\label{app:related}

\descr{Membership/Attribute Inference Attacks.}
As mentioned, generative models trained without robust DP guarantees can memorize and/or overfit to individual train data points~\citep{webster2019detecting} and overall might be vulnerable to membership and attribute inference attacks---for short, MIAs and AIAs, respectively. %
Hayes et al.~\citep{hayes2019logan} present the first MIA against GANs and Variational Autoencoders (VAEs).
They do so by training a shadow discriminator (the original one in a white-box setting and a purpose-built one in a black-box setting) to output the data with the highest confidence as the train data.
Hilprecht et al.~\citep{hilprecht2019monte} study both MIAs and reconstruction attacks based on Monte Carlo integration against GANs and VAEs. %
Chen et al.~\citep{chen2020gan} present a taxonomy of MIAs and a novel generic attack against GANs; then, Zhu et al.~\citep{zhu2023data} introduce the first MIA against diffusion models.
Stadler et al.~\citep{stadler2022synthetic} present a systematic evaluation of MIAs and AIAs in the context of tabular synthetic data and show that outliers are vulnerable when model are trained without DP guarantees.
Annamalai et al.~\citep{annamalai2023linear} introduce a new attribute inference attack based on linear reconstruction.
Overall, as discussed in Section~\ref{subsec:mia}, unlike state-of-the-art MIAs/AIAs, \emph{DifferenceAttack} requires no reference data and relies on a single pre-trained model.

\descr{Reconstruction Attacks in Databases.}
Dinur and Nissim~\citep{dinur2003revealing} present the first reconstruction attack where the adversary can theoretically reconstruct records from a database consisting of $n$ entries by sending count queries and solving a linear program.
The adversary can make at most $n$ queries, while the answers must be highly accurate. %
Follow-up studies~\citep{dwork2007price, dwork2008new} generalize and improve on the results by relaxing some of the assumptions and achieving better reconstruction rates.
While these attacks are of a theoretical nature, they have %
contributed to the rigorous definition of DP~\citep{dwork2006calibrating}.
Also, reconstruction attacks on aggregate statistics contributed to the US Census Bureau's deployment of DP for the 2020 Census~\citep{garfinkel2019understanding}.
More recently, Dick et al.~\citep{dick2023confidence} reconstruct private records based on aggregate query statistics and public distributions while reliably ranking them.

\descr{Reconstruction Attacks in ML.}
Reconstruction attacks are %
sometimes referred to as model inversion attacks~\citep{fredrikson2014privacy, fredrikson2015model, yeom2018privacy}.
Zhu et al.~\citep{zhu2019deep} demonstrate how an adversary with access to model gradients can efficiently use them to reconstruct train records; the recovery is pixel-wise accurate for images and token-wise matching for text.
In online and federated learning settings, attackers can infer train data points or their labels from the intermediate gradients during training~\citep{wang2019beyond, geiping2020inverting, salem2020updates, zanella2020analyzing}.
This assumes one can observe the gradient updates of the target model multiple times, whereas we have black-box access to a single trained model.

Train data extraction attacks, which could also be considered reconstruction, have been proposed in various contexts, e.g., (generative) large language models~\citep{carlini2019secret, carlini2021extracting} and diffusion models~\citep{carlini2023extracting}.
These usually assume some auxiliary knowledge (e.g., the presence of the target in the train data or, similarly to attribute inference, a subset of the target's attributes) and query the model multiple times to exploit their memorization vulnerability.
In other words, they exploit the tendency of large models to memorize and reproduce the train data at generation.
Finally, Balle et al.~\citep{balle2022reconstructing} and Haim et al.~\citep{haim2022reconstructing} propose reconstruction attacks against discriminative models in which the adversary either has access to all data points but one or to the trained weights and reconstruct the remaining one/several train records.
In contrast, \emph{ReconSyn} does not assume any knowledge about the model/data and also works when the model has no memorization capability.

\section{Discussion and Conclusion}
\label{sec:conclusion}
In this section, we recap our findings, consider potential (but ineffective) countermeasures against our attacks, and conclude the paper with a few additional observations.

\subsection{Summary}
Our work showcases the fundamental limitations of reasoning about synthetic data privacy purely through empirical evaluation and similarity-based privacy metrics (SBPMs), demonstrating the unreliability and inconsistency of these privacy heuristics used by the leading providers. %
The effectiveness of our novel reconstruction attack, \emph{ReconSyn}, %
consistently shows that privacy protections can be meaningless even if all privacy tests pass.
Over different models and datasets, we can completely reconstruct -- and thus single out/link to -- most outliers, failing two of the required GDPR protections.
As a result, synthetic data with privacy guaranteed through SBPMs cannot be considered anonymous.

In a way, one could compare SBPMs to the inadequate privacy guarantees offered by Diffix-like anonymization systems~\citep{knockdiffix,gadotti2019signal, cohen2020linear}. %
Although the functionalities are different (Diffix provides answers to queries), they both allow for a potentially unbounded number of queries while not implementing robust privacy mechanisms like DP ultimately leading to severe privacy violations.

We believe our findings will be useful to practitioners deploying solutions that require processing and releasing sensitive data, and help policymakers create standards and best practices for privacy-preserving synthetic data adoption.

\descr{Responsible Disclosure.}
Our main goal is not to attack live systems or access any personal data but to demonstrate the need for formal notions of privacy when rolling out systems in production.
We shared our work with the two main synthetic data companies using SBPMs, providing them with 90 days for a response while keeping the paper confidential.
Since then, one company stopped offering privacy filters alongside DP model training while the other started offering DP model support alongside SBPMs and updated its marketing materials.

\descr{Competing Interests.}
This work was done while GG was with Hazy, a synthetic data provider.
EDC is an academic researcher with no competing interests in this work.

\subsection{Potential Countermeasures?}
\label{subsec:mit}
We now discuss the effectiveness of possible mitigations such as training the generative models with DP guarantees, limiting the number of calls, DP-fying the metrics, or preventing overfitting/memorization. %

\subsubsection{DP Training}
Training generative models with DP guarantees is the established approach to mitigate attacks against synthetic data.\footnote{Other possible defense strategies, such as regularization, dropout, or knowledge/model distillation, have largely proven ineffective~\cite{papernot2018sok}.}
However, combining DP training with access to the metrics, as done by a few companies (cf.~Section~\ref{subsec:solution}), does not provide meaningful defenses.
To confirm this, we perform an experimental evaluation with four DP models that rely on different mechanisms -- namely, PrivBayes (Laplace), MST (Gaussian), DPGAN (DP-SGD), and PATE-GAN (PATE) -- with privacy budgets in \{0.1, 1, $\infty$\} on the \emph{Adult Small} dataset.
We keep $\delta$ constant to $1/n$.
We also use the non-DP model CTGAN for comparison.
Again, we use \emph{SampleAttack} (1,000 rounds) on all models and \emph{SearchAttack} (up to 1 column) where the former fails to achieve at least 95\% reconstruction success.

\begin{figure}[t!]
	\vspace{-0.3cm}
	\centering
	\begin{subfigure}{0.81\linewidth}
		\includegraphics{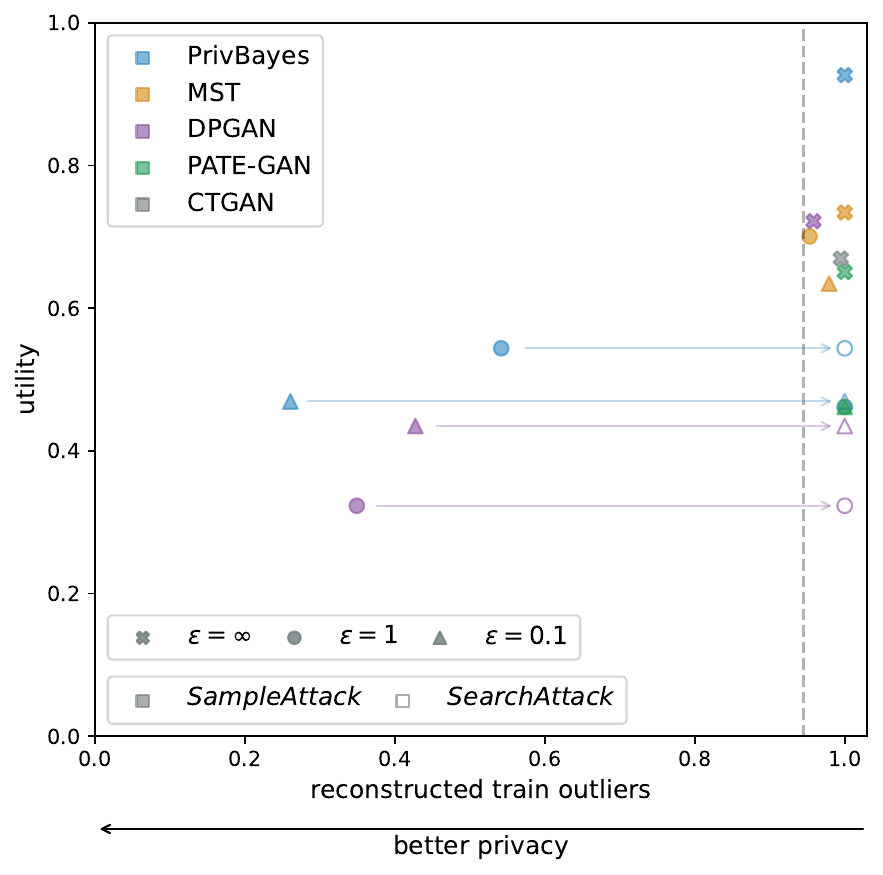}
	\end{subfigure}
	\caption{Synthetic data utility vs.~proportion of train outliers reconstructed by \emph{ReconSyn} on \emph{Adult Small}. (Left-to-right arrows link the performance of using only {\em SampleAttack} to also using {\em SearchAttack} in the same settings.) }
	\label{fig:privacy_utility}
	\vspace{-0.2cm}
\end{figure}

We report the resulting privacy-utility tradeoffs in Figure~\ref{fig:privacy_utility}.
As in prior work~\citep{tao2022benchmarking}, we measure utility through similarity, %
reporting the average difference between all 1-way marginals and 2-way mutual information scores (normalized in [0,1]) across train and synthetic data.
As expected, applying DP reduces utility across the board.
The utility of MST drops less than PrivBayes, and similarly, PATE-GAN less than DPGAN, due to the different underlying DP mechanisms, as pointed out in~\citep{ganev2024graphical}.

We measure privacy in terms of the proportion of train outliers reconstructed by \emph{ReconSyn}.
Regardless of the model, privacy budget, or utility, we successfully recover more than 95\% of its targets (to help visualize this, we add a dashed vertical line to Figure~\ref{fig:privacy_utility}).
Applying DP to higher utility models -- i.e., MST and PATE-GAN -- does not even defend against \emph{SampleAttack}, while it yields a big drop in utility for PrivBayes and DPGAN.
Nonetheless, \emph{SearchAttack} also recovers all train outliers against these two models.

Training models with DP guarantees does not help because the leakage comes from the privacy metrics; since they require access to the train data and are deterministic (as discussed in I4 in Section~\ref{sec:issues}), they break the end-to-end DP pipeline.
Any additional privacy mechanism applied on top of the metrics is unlikely to mitigate the problem, as the analysis of the privacy guarantees of the whole system needs careful consideration~\cite{debenedetti2024privacy}.

\subsubsection{DP-fying the Metrics}
Theoretically, an intuitive countermeasure would be applying DP mechanisms to (``DP-fying'') the metrics while still running statistical pass/fail tests.
For instance, the provider could select the closest record using the Exponential mechanism~\citep{dwork2006our}. %
However, this would not offer a robust solution either.
DP-fying the metrics further increases the privacy budget spent with each generation run, as the metrics must be applied every time.
Eventually, the allocated DP budget will be reached, preventing further data generation.
This contradicts one of synthetic data's main purported advantages, i.e., the ability to generate unlimited data.

\subsubsection{Disabling Access to the Metrics}
Since the leakage comes from the metrics, one might think that disabling access to the metrics scores altogether would easily solve the problem.
However, doing so would compromise the provider's transparency and explainability promises; %
the users would need to rely solely on the provider’s assurance that the generated synthetic data meets the required metrics above a given threshold.
Users cannot verify compliance claims without tangible proof of passing these tests, undermining another one of the provider’s primary selling points.

Additionally, if the provider relies on the statistical tests and only provides a success/fail flag (without scores), a strategic adversary %
could, in principle, still extract sensitive information, e.g., reconstructing approximate matches with at least 95\% similarity to the nearest train record.

\subsubsection{Limiting Number of Calls}
Alternatively, the provider could set global or per-user limits on the number of API calls.
However, this would also undermine the promise of unlimited data.
Moreover, detecting an unusually high volume of calls in real time could be challenging, as generating large amounts of data to simulate various edge cases is often part of normal testing activities.
More importantly, detection may be unfeasible altogether, as synthetic data platforms are frequently deployed on clients' premises or private clouds~\citep{gretel2024deploy, tonic2024deploy, mostly2024deploy, hazy2024deploy}, where providers lack direct access to monitor and mitigate such risks.
We also demonstrated that, in some settings, even a handful of calls can enable very accurate attacks, leaving to future work to optimize the number of calls to obtain optimal trade-offs.
Incidentally, as discussed in Section~\ref{sec:model}, providers allow users to generate large numbers of queries, which is also a standard assumption in previous related attacks~\citep{carlini2024stealing, nasr2023scalable}.

\subsubsection{Preventing Overfitting/Memorization}
Privacy attacks against synthetic data are generally correlated with generative models memorizing and/or overfitting
~\cite{carlini2019secret, webster2019detecting, hayes2019logan}.
However, we argue that reconstruction attacks would still be possible in the context of SBPM-based techniques even if one could somehow guarantee that the underlying models do not overfit or memorize the data.

To validate this hypothesis, we show that \emph{ReconSyn} is successful even against models with restricted capabilities, such as Independent and Random, on \emph{Adult Small}.
Since these models cannot accurately represent the data, the adversary cannot locate the clusters with outliers through \emph{OutliersLocator}.
Instead, we set their goal to reconstruct {\em any} train data points.
Keeping the same settings, we launch \emph{ReconSyn}'s first subattack, \emph{SampleAttack}, for 1,000 rounds.

The adversary recovers around 79\% of the train data against both models.
Unsurprisingly, the recovery rate on Random is much slower than Independent, i.e., more rounds are needed to achieve comparable results.
Incidentally, in both cases, the adversary reconstructs all 192 train outliers.
This could be due to the small data cardinality and randomness component, as both models have a higher chance of generating data points with low probability compared to the five main models, which learn to generate realistic data better.
If the adversary successfully reconstructs a large proportion of the train data points, they could use them to fit \emph{OutliersLocator} and locate the outliers as a last step.

\subsection{Further Considerations}
\label{sec:further}

\noindent{\bf Remarks on DP.}
With end-to-end DP pipelines, privacy becomes an attribute of the process~\citep{trask2020beyond}, and, as per the post-processing property, any synthetic data sample is also DP.
DP also provably bounds the risks of singling out predicates~\citep{cohen2020towards} and of linkability and inference attacks~\citep{giomi2022unified}. %

Having said that, training generative models with DP also poses some challenges, which may explain why several companies opt for heuristics.
Since there is no one-model-fits-all for all use cases~\citep{jordon2022synthetic}, it must be determined whether DP addresses the right threat.
For instance, %
if the dataset contains proprietary secrets, e.g., company-specific terms/names/locations, these may still be exposed in DP synthetic data unless%
~\citep{nhs2021ae}.
Moreover, selecting the optimal combination of the generative model and DP mechanism is not straightforward, as it depends on factors such as the privacy budget, downstream task complexity, dataset dimensionality, imbalance, and domain~\citep{ganev2024graphical}.
Thus, determining the right privacy budget is highly context-specific~\citep{hsu2014differential}.

Though a key strength, DP's {\em worst-case} guarantees often reduce utility in a disproportionate way across different records, particularly for outliers~\citep{stadler2022synthetic, kulynych2023arbitrary} and underrepresented classes/subgroups~\citep{bagdasaryan2019differential, ganev2022robin}.
For some use cases, any significant drop in utility may not be an option.
Combining generative models and DP mechanisms, unfortunately, could result in unpredictable data; e.g., it might not be clear what signals/trends will be preserved~\citep{stadler2022synthetic}, a fundamental requirement of usable privacy mechanisms~\citep{edps2018preliminary}.
Overall, implementing DP in practice and effectively communicating its properties is known to be non-trivial~\citep{cummings2021need, houssiau2022on}.

\descr{Empirical Evaluations.}
While our work demonstrates that SBPMs should not be used as attempts to guarantee privacy, empirical evaluations and attacks should not be disregarded altogether.
They can be valuable tools to detect flaws, errors, or bugs in algorithms and implementations, aiding in DP auditing~\citep{jagielski2020auditing, tramer2022debugging, nasr2023tight}, and enhancing the interpretability of theoretical privacy protections~\citep{houssiau2022framework, houssiau2022tapas}.

\descr{Limitations and Future Work.}
\emph{ReconSyn} successfully reconstructs most outliers in various settings, yet it could still be optimized.
In future work, we plan to investigate the minimum number of records the adversary needs to generate %
as well as the effect of preventing the adversary from augmenting the generated synthetic data.
Follow-up research could also study the impact of privacy metrics using continuous distances, %
which could make the calculations more precise and yield better search algorithms. %

\descr{Acknowledgments.}
We are grateful to the IEEE S\&P shepherd and reviewers for their valuable feedback, which helped us to improve our paper.
We also thank Theresa Stadler and Yves-Alexandre de Montjoye for their helpful comments.

{\footnotesize
\bibliographystyle{abbrv}

}

\appendix

\section{Reconstruction of All Train Data}
\label{app:all}

\begin{table}[t!]
	\footnotesize
	\centering
	\begin{tabular}{l|rr|r}
		\toprule
			\textbf{Model}				& \multicolumn{2}{c|}{Any Train Records}			& Train Outliers	\\
														& \emph{Sample}						& \emph{Search}			& \emph{Search}		\\
		\midrule
			\textbf{PrivBayes}		&	0.20										& 0.98							& 0.95						\\
			\textbf{MST}					&	0.33										& 0.91							& 0.74						\\
			\textbf{DPGAN}				&	0.04										& 0.85							& 0.51						\\
			\textbf{PATE-GAN}			&	0.07										& 0.85							& 0.50						\\
			\textbf{CTGAN}				&	0.06										& 0.84							& 0.48						\\
	 \bottomrule
	\end{tabular}
	\vspace{2pt}
	\caption{Reconstruction rate of any records by \emph{ReconSyn}, \emph{Adult}.}
	\label{tab:any_train}
\end{table}

To validate the hypothesis that reconstructing outliers is inherently more difficult, %
we conduct an experiment aimed at recovering {\em any} train records from \emph{Adult}.
We impose stricter constraints than those in Section~\ref{sec:rec}, i.e., we limit \emph{SampleAttack} to only 250 rounds, down from 1,000, and allow \emph{SearchAttack} to trace just 3 distances instead of 4.
The results are summarized in Table~\ref{tab:any_train}.

We observe that even under these limited settings, \emph{ReconSyn} reconstructs a significant proportion of the train data.
Specifically, using only a fraction of the computations, the average recovery rate of any train data is 0.89\%, which exceeds the average recovery rate of outliers for \emph{Adult} (average of 0.85\%) as presented in Section~\ref{sec:rec}.
Overall, this should not be a surprise, given that outliers are less likely to be generated.
The rightmost column in Table~\ref{tab:any_train} further supports this, showing a substantial drop in recovering the outliers for all models, with PrivBayes being the only exception.

\section{Datasets}
\label{app:data}
We describe the datasets used for our evaluation and our criteria for defining outliers;
see Table~\ref{tab:datasets}. %
We construct two controllable datasets (\emph{2d Gauss} and \emph{25d Gauss}) based on the normal distribution and use three `standard' datasets after some sampling/pre-processing (\emph{Adult}, \emph{Census}, and \emph{MNIST}).

\descr{\emph{2d Gauss}.}
We sample 2,000 points from a standard bivariate normal distribution with zero correlation.
We consider all train points beyond the blue circle displayed in Figure~\ref{fig:2d_gauss_fixed} (centered at 0, radius 2.15) outliers, or about 10\%.

\descr{\emph{25d Gauss}.}
This is similar to \emph{2d Gauss} but extended to 25 dimensions (again, standard normal distribution).
Note that we do not reconstruct outliers for this dataset and we only use it in a counter-example in Appendix~\ref{app:examples} (CE6).

\begin{figure}[t]
\centering
		\begin{subfigure}{\mywidth\linewidth}
			\includegraphics{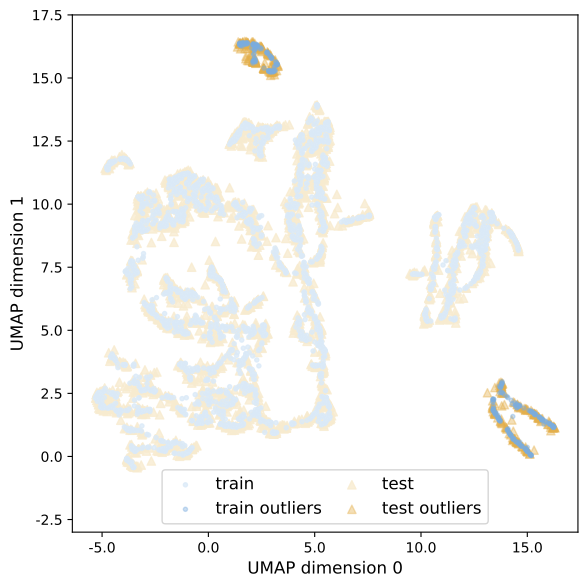}
			\caption{2d UMAP, \emph{Adult Small}}
			\label{fig:adult_small}\smallskip
		\end{subfigure}
		\hfill
		\begin{subfigure}{\mywidth\linewidth}
			\includegraphics{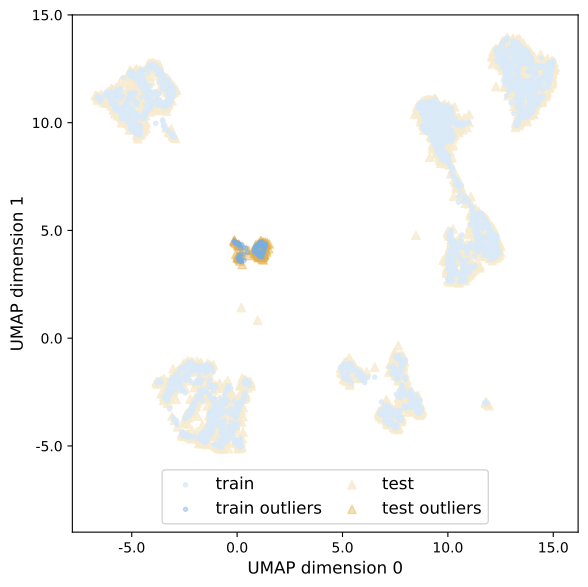}
			\caption{2d UMAP, \emph{Adult}}
			\label{fig:adult}
		\end{subfigure}
		\begin{subfigure}{\mywidth\linewidth}
			\includegraphics{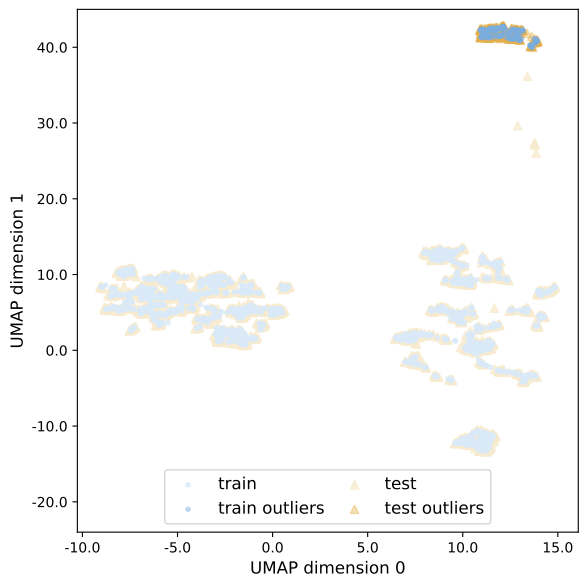}
			\caption{2d UMAP, \emph{Census}}
			\label{fig:census}
		\end{subfigure}
				\hfill
		\begin{subfigure}{\mywidth\linewidth}
			\includegraphics{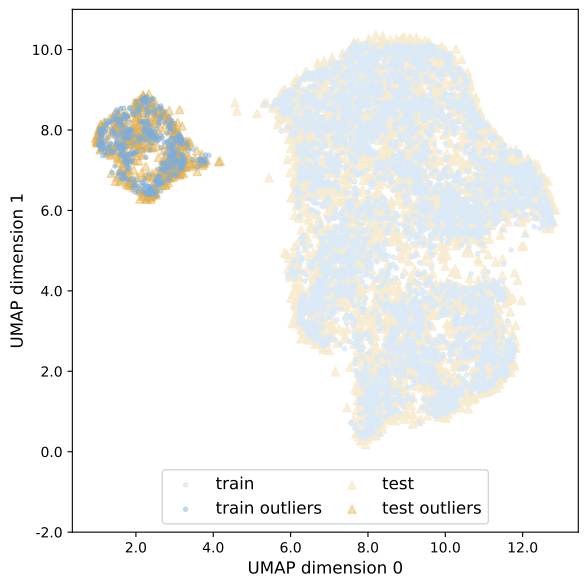}
			\caption{2d UMAP, \emph{MNIST}}
			\label{fig:mnist}
		\end{subfigure}\\[1ex]
	\vspace{0.1cm}
	\caption{Train and test data, 2d UMAP reduced.} %
	\label{fig:datasets}
\end{figure}

\descr{\emph{Adult}.}
We use two versions of the Adult dataset~\citep{dua2017adult}.
We randomly sample 6,000 data points and refer to this dataset as \emph{Adult}.
Then, we further simplify this dataset by selecting six columns (`age,' `education,' `marital status,' `relationship,' `sex,' and `income'), and denote the resulting dataset as \emph{Adult Small}.
For both, we fit a Gaussian Mixture model with 10 clusters; for the former, we select the smallest cluster to be outliers, while for the latter, the smallest two (2d UMAP reduction shown in Figure~\ref{fig:adult_small} and~\ref{fig:adult}).

\descr{\emph{Census}.}
We randomly sample 10,000 data points from the Census dataset~\citep{dua2017adult}.
To determine the outliers, we fit a Gaussian Mixture model with 4 clusters and select the smallest one (2d UMAP reduction is plotted in Figure~\ref{fig:census}).

\begin{table}[t]
	\footnotesize
	\centering
	\setlength{\tabcolsep}{5pt}
	\begin{tabular}{lrrrr}
		\toprule
			\bf Dataset	 				 & \bf Cardinality	& \bf \#Cols 			& \bf \#Records		& \bf \#Outliers	\\
		\midrule
			{\em 2d Gauss}	 	 	 & $10^{5}$					& 2								& 2,000						& 108							\\
			{\em 25d Gauss}			 & $10^{31}$				& 25							& 2,000						& N/A							\\
		 	{\em Adult Small}	 	 & $10^{5}$					& 6								& 6,000 					& 192							\\
			{\em Adult}		 		 	 & $10^{15}$				& 14				 			& 6,000 					& 116							\\
			{\em Census}			 	 & $10^{43}$				& 41				 			& 10,000 					& 193							\\
		 	{\em MNIST}				 	 & $10^{78}$				& 65							& 10,000 					& 488							\\
		\bottomrule
	\end{tabular}
		\vspace{2pt}
	\caption{Summary of the six tabular datasets used throughout our experiments, reporting their cardinality, the number of columns and records they include, and that of train outliers.}
	\label{tab:datasets}
\end{table}

\descr{\emph{MNIST}.}
We sample 9,000 data points from the digits 3, 5, 8, and 9 %
from the \emph{MNIST}~\citep{lecun2010mnist} dataset as well as 1,000 from 0 and treat them as outliers (2d UMAP reduction displayed in Figure~\ref{fig:mnist}).
To simplify the dataset, we downscale the images to 8x8 pixels and discretize all pixels to 16 bins.

\descr{Outliers Definition.}
While various definitions of outliers exist in literature~\citep{carlini2019distribution, meeus2023achilles}, we define underrepresented data regions and outliers to capture various scenarios, aiming to intuitively select roughly 10\% of the train data (to serve as targets, as noted in Table~\ref{tab:datasets}).
For \emph{2d Gauss}, we identify outliers as points lying beyond a certain distance from the center.
In \emph{Adult Small}, \emph{Adult}, and \emph{Census}, outliers are the smallest clusters determined by a Gaussian Mixture model.
For \emph{MNIST}, the digit 0 is deliberately underrepresented.
We apply the same strategy/model to the test data.
UMAP's distance-preserving feature allows these outlier identification strategies to be visually verified in Figure~\ref{fig:adult_small}--\ref{fig:mnist}. %

\descr{Comparison with Previous Studies.}
Our experiments are conducted on more datasets with higher dimensionality and complexity than previous attacks against tabular synthetic data~\cite{stadler2022synthetic, houssiau2022framework, houssiau2022tapas, van2023membership, annamalai2023linear, meeus2023achilles, guepin2023synthetic}.
These studies evaluate membership/attribute inference attacks versus, at most, two datasets with a dimensionality of no more than 35.
By contrast, we run a more challenging task (i.e., reconstruction attacks) on five datasets.
Moreover, some of the datasets included in our evaluation are more complex and have higher dimensions, namely, 41 (Census) and 65 (MNIST).

\begin{figure*}[t!]
	\centering
	\begin{subfigure}{\mywidth\linewidth}
		\includegraphics{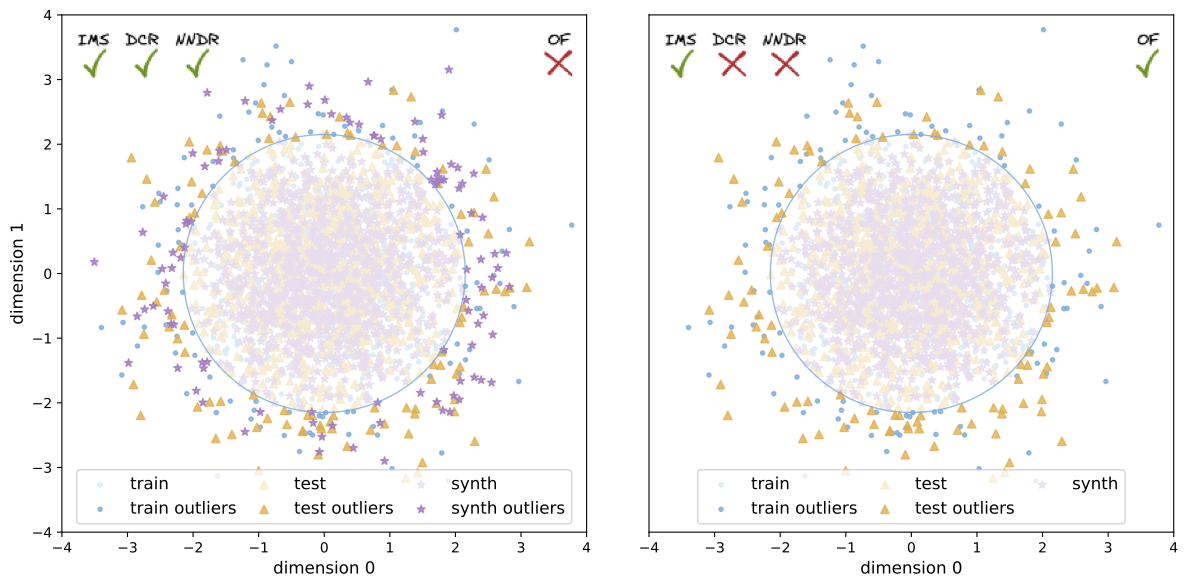}
		\caption{From PPP to PFF}
		\label{fig:wonky_filter_0}
	\end{subfigure}
	\begin{subfigure}{\mywidth\textwidth}
		\includegraphics{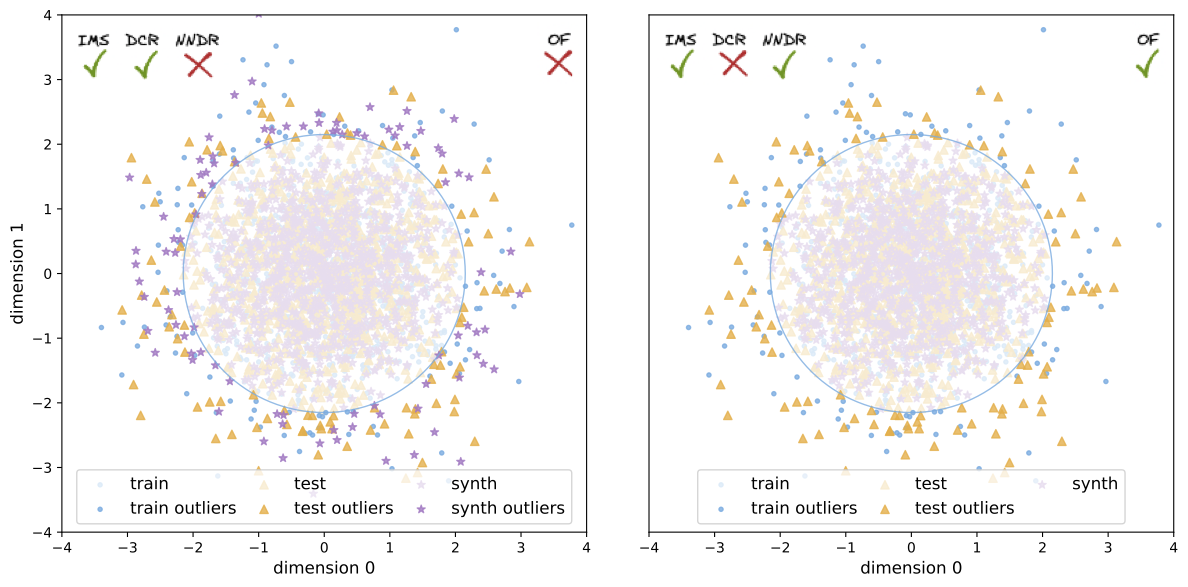}
		\caption{From PPF to PFP}
		\label{fig:wonky_filter_1}
	\end{subfigure}\\[1ex]
	\caption{Examples of privacy metrics unreliability before/after applying OF, \emph{2d Gauss}.}
	\label{fig:wonky}
\end{figure*}

\section{Additional Fundamental Issues of SBPMs}
\label{app:issues}
We discuss three more fundamental issues with SBPMs.

\descr{I7: Incorrect Interpretation.}
From a statistical theory point of view, the results of the privacy metrics pass/fail tests can be misinterpreted.
Assuming a good statistical test, the null hypothesis ($H_0$) %
is ``privacy is preserved,'' while the alternative hypothesis ($H_A$) becomes ``privacy is not preserved.''
When the observed data supports $H_A$, %
we can reject $H_0$ and claim that we have detected privacy violations.
However, when the observed data does not allow us to reject $H_0$, this simply means that we fail to reject $H_0$.
We cannot claim that $H_0$ is accepted or ``privacy is preserved.''

\descr{I8: Risk Underestimation.}
Most implementations of the privacy metrics %
require discretizing numerical columns and using Hamming distance to compute the similarity between data points.
Unfortunately, the calculations become imprecise, and the privacy protections are overstated compared to, for example, using continuous data and Euclidean distance.

\descr{I9: Data Access.}
Due to the sensitive nature of the data, it is imperative to train the generative model within the secure environment the data resides.
Once trained, the model cannot be exported since the privacy metrics require access to the train data for each generation run.
This prompts a challenge where accessing the secure environment becomes necessary for every synthesized data.

\begin{table}[t]
	\footnotesize
	\centering
	\setlength{\tabcolsep}{4pt}
	\begin{tabular}{rrrrrr}
		\toprule
			\bf \#Cols	 	& \bf \#Outliers	& \bf \#Synth 		 & \bf \#Bins 	& \bf \#Bins 		& \bf	Attack		\\
										&									& \bf Datasets		 & \bf Outliers & \bf Outliers  & \bf Success		\\
										& 								&	\bf (* $10^{3}$) & \bf Total    & \bf Empty		  & 							\\
		\midrule
			{\em 2}	 	 	 	& 108							& 8								 & 442					& 108						& 1.00					\\
			{\em 3}			 	& 118							& 4,059						 & 12,632				& 140						& 0.84					\\
		 	{\em 4}	 	 		& 110							& 3,520						 & 340,475			& 26,236				& 0.01					\\
			{\em 5}		 		& 95							& 757							 & 8,915,441		& 4,573,841			& 2.08*$10^{-5}$ \\
		\bottomrule
	\end{tabular}
	\vspace{2pt}

	\caption{Similarity Filter and reconstruction success rate (12~hour limit), \emph{Gauss} with different dimensions.}
	\label{tab:recon_sf}
\end{table}

\section{Additional SBPMs Counter-Examples}
\label{app:examples}
This section presents three additional counter-examples highlighting the inconsistency of the privacy metrics/filters.
For the first two use \emph{2d Gauss}, and the last uses \emph{25d Gauss}.

\descr{\em CE3.~SF \& Reconstruction (continued).}
We expand CE3 by extending \emph{Gauss} to higher dimensions and present the results in Table~\ref{tab:recon_sf}.
While we successfully isolate all outliers for two dimensions and 84\% for three, the number of bins potentially containing outliers grows exponentially, reaching nearly 350,000 for four dimensions and 9,000,000 for five.
Consequently, the attack slows down significantly; within 12 hours of runtime, we can only `visit' a small fraction of these bins, rendering the attack ineffective.

Conversely, targeting a single data record, as in standard membership inference attacks, makes the attack more effective.
Suppose the target lies at the origin (the zero vector) with no training records nearby.
Without conditional data generation, i.e., we naively sample from an oracle, we can create a `hole' around the target for up to 8 dimensions in less than 3 hours.
Beyond this, the neighborhood's dimensionality grows too large.
However, with conditional data generation, we can construct the target's neighborhood one column at a time, conditioning on zeros for the remaining columns.
This reduces complexity from exponential to linear, enabling us to encircle the target even in dimensions beyond 100 within seconds.

\descr{\em CE4.~Metrics Inconsistency.}
Suppose we rely on the oracle to sample 1,000 new datasets.
A good privacy metric should reflect that the oracle perfectly preserves the train data privacy by reporting a high privacy score.
Only on 274 occasions (out of 1,000), all the privacy tests pass.
This demonstrates that metrics and empirical approaches measuring the privacy of a single synthetic dataset completely fail to capture the generating process.

Moreover, the proportion of times when the individual metrics IMS, DCR, and NNDR pass is 1, 0.48, and 0.38, respectively, which is widely inconsistent.
Even though the synthetic datasets were sampled from a fixed distribution, which can be thought of as a generator, DCR and NNDR report random results that are not close to 0 or 1.
In practice, this means that even if the generative model captures the underlying generating process well, without overfitting or memorizing the train data, the pass/fail decision depends on a specific sample, is noisy, and cannot be trusted.

Alternatively, fixing a synthetic dataset and randomly splitting the sensitive data 50/50 into train/test sets still leads to inconsistencies.
Out of 1,000 repetitions, only 380 instances pass all three tests.
This highlights the inherent randomness in the train/test split, which incorporates instability into the evaluation process.

\descr{\em CE5.~OF \& Metrics Inconsistency.}
We also examine how applying the OF, which is supposed to improve privacy, affects it according to the metrics.
Again, we rely on the oracle to draw synthetic data samples.
On the left plot of Figure~\ref{fig:wonky_filter_0}, we see that the synthetic data passes all 3 tests, while on the right plot, when the outliers are filtered out, both DCR and NNDR fail.
Even more surprisingly, Figure~\ref{fig:wonky_filter_1} shows that removing the outliers can cause a previously failing test to pass (NNDR) and vice versa (DCR).
These examples further prove the untrustworthiness and inconsistency of the privacy metrics and filters.
One could observe all possible combinations of DCR and NNDR passing/failing.

\begin{figure}
	\centering
	\includegraphics[width=0.99\columnwidth]{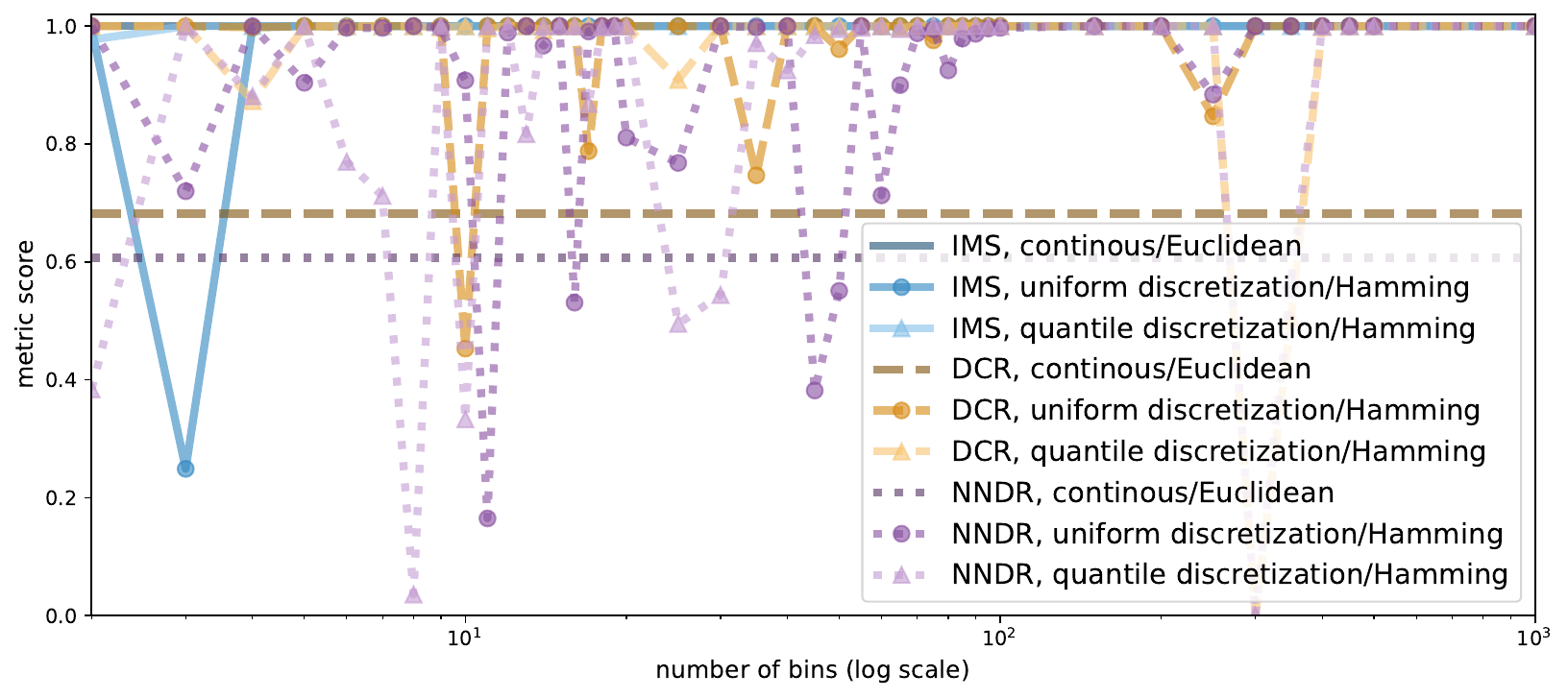}
	\caption{Discretization effect on privacy metrics, \emph{25d~Gauss}.}
	\label{fig:wonky_n_bins}
\end{figure}

\descr{\em CE6.~Discretization Inconsistency.}
Finally, in Figure~\ref{fig:wonky_n_bins}, we measure the effect of discretizing data.
We test two discretization strategies, uniform and quantile, while varying the number of bins ($n_{bins}$) from 2 to 1,000.
For discrete data, we use Hamming distance, while for continuous data, Euclidean.
We test on \emph{25d Gauss}, use an oracle to sample 1,000 synthetic datasets, and report averages.

First, the continuous results show that both DCR and NNDR report average values roughly around 0.65 (again failing to capture the generating process, similarly to previous examples).
Second, discretizing the data and using Hamming distance greatly overestimates privacy -- DCR and NNDR have scores of around 1, except for some randomly looking drops.
Incidentally, the metrics report approximately correct results but for the wrong reasons.
They overestimate privacy due to discretization but fail to capture the generating process.
Last, varying the discretization strategy and the number of bins does not help any metrics become more accurate (closer to the continuous baseline).

Regarding the randomly looking score drops, we have two hypotheses.
First, most drops (except 4) appear when the number of bins is below 100.
This could be because with larger numbers of bins, we discretize the space into exponentially many hypercubes ($n_{bins}^{25}$ since there are 25 columns), and it becomes increasingly unlikely to get two numbers in the same hypercube even if their actual Euclidean distance is small.
Second, the most drops occur when using the uniform strategy (at least for IMS and DCR).
This could be because of the nature of the dataset (i.e., normal distribution).
For uniform discretization, naturally, the bins in the middle will contain most of the data records (while there will be very few in the tails), increasing the chance of a match.

\end{document}